# Informe Técnico / Technical Report

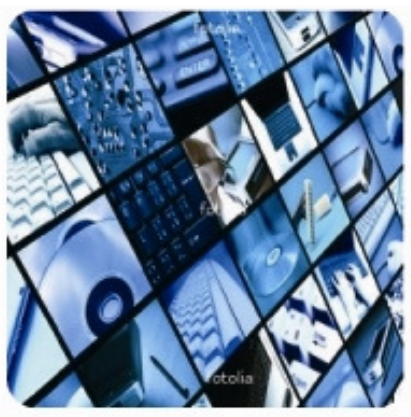
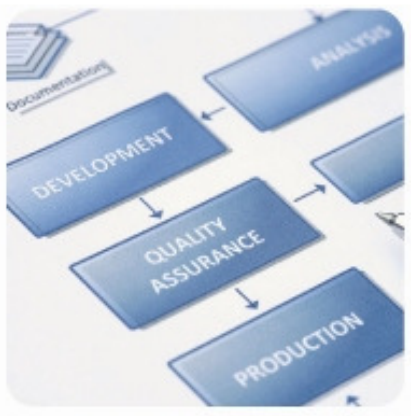

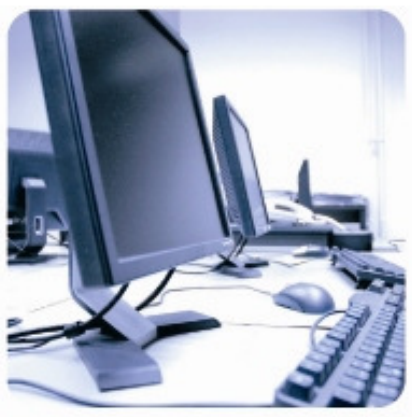
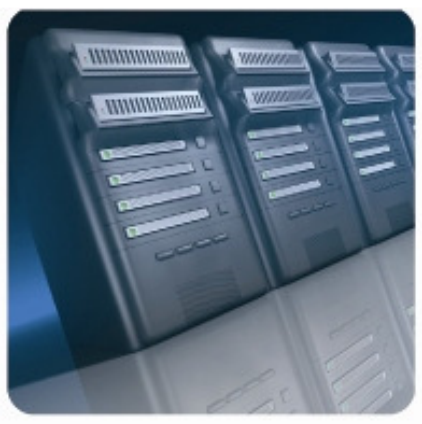
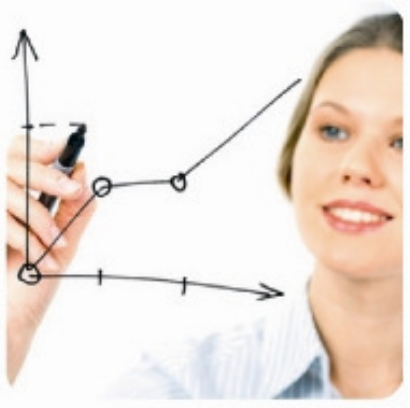

## Integration of Communication Analysis and the OO Method: Rules for the manual derivation of the Conceptual Model

**Sergio España, Arturo González, Óscar Pastor, Marcela Ruiz**

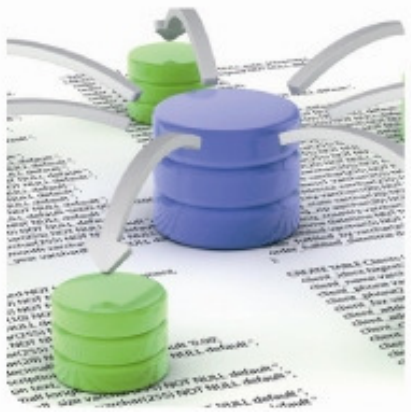
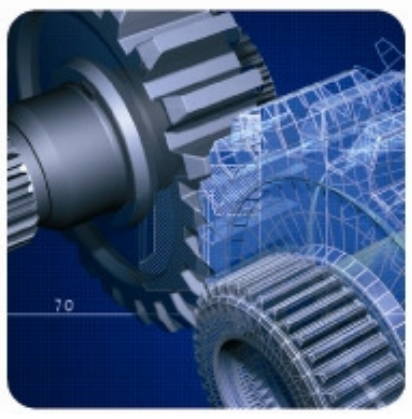
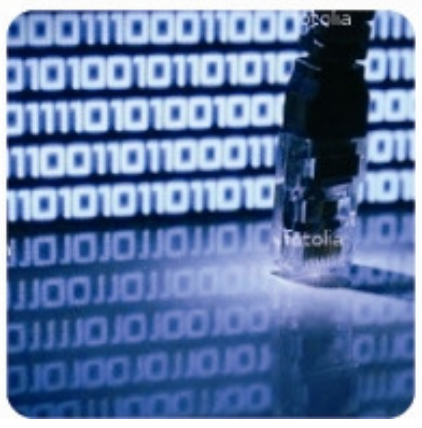
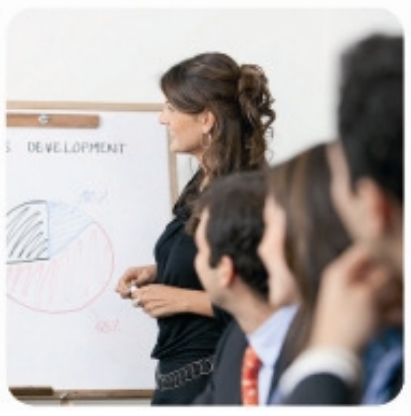
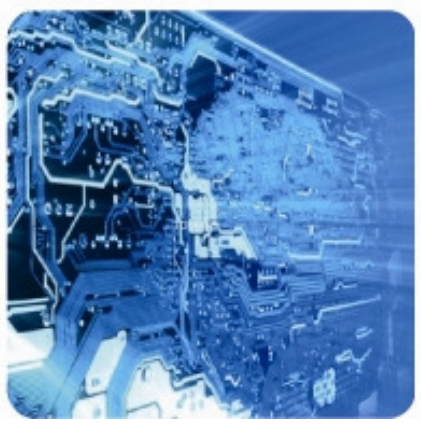

| | | | | | |
|---|---|---|---|---|---|
| Ref. #: | ProS-TR-2011-04 | | | | |
| Title: | Integration of Communication Analysis and the OO-Method: Rules for the manual derivation of the Conceptual Model | | | | |
| Author (s): | Sergio España, Arturo González, Óscar Pastor, Marcela Ruiz | | | | |
| Corresponding author (s): | Sergio España sergio.espana@pros.upv.es<br>Marcela Ruiz lruiz@pros.upv.es | | | | |
| Document version number: | 2.0 | Final version: | NO | Pages: | 38 |
| Release date: | May 2011 | | | | |
| Key words: | Information systems, analysis, design, requirements engineering, conceptual modelling, model-driven development, model derivation, Message Structures, Communication Analysis, OO-Method, *OLIVANOVA* | | | | |

Universidad Politécnica de Valencia · Camino de Vera s/n · Edificio 1F · 46022 Valencia Spain ·
T. +34 96 387 70 07 Ext. 83530 · M. + 34 619 543 623 · F. +34 96 387 73 59 · info@pros.upv.es · www.pros.upv.es

# TABLE OF CONTENTS

# 1. INTRODUCTION

## 1.1. Motivation of this work

This document presents a handbook that describes the derivation of OO-Method conceptual models from Communication Analysis requirements models. The main objective is to describe a set of rules that guide the derivation process; that is, a model transformation intended for analysts is specified. Also, the rules are illustrated using simple examples.

A more detailed application of the derivation guidelines is described in detail in the SuperStationery Co. case [1], which presents the business process model and requirements model that specify the organisational work practice of a company that sells office material, as well as how the corresponding conceptual model is derived.

The derivation guidelines presented in the following chapters have been published in conference papers [2][3]. Therefore, if you intend to cite this technical report, please consider citing also/instead the following papers:

A. González, S. España, M. Ruiz, and Ó. Pastor, "Systematic derivation of class diagrams from communication-oriented business process models." In: 12th Working Conference on Business Process Modeling, Development, and Support (BPMDS'11), London, United Kingdom, 2011.

S. España, M. Ruiz, A. González, and O. Pastor, "Systematic derivation of state machines from communication-oriented business process models Derivation of the dynamic view of conceptual models in an industrial model-driven development method." In: Fifth IEEE International Conference on Research Challenges in Information Science (RCIS'11), Guadeloupe, France, 2011.

## 1.2. Scope of this document

This document does not describe Communication Analysis and the OO-Method in full detail. For further information about these methods please refer to [González, España et al. 2008; España, González et al. 2009; España, Condori-Fernández et al. 2010; González, Ruiz et al. 2011] and [Pastor and Molina 2007], respectively.

# 2. DERIVATION OF THE OBJECT MODEL

## 2.1. Derivation Strategy

We first present a graphic overview of the derivation strategy (see Figure 1); both the Communicative Event Diagram and the Event Specification Templates contain relevant information for the derivation of the Object Model.

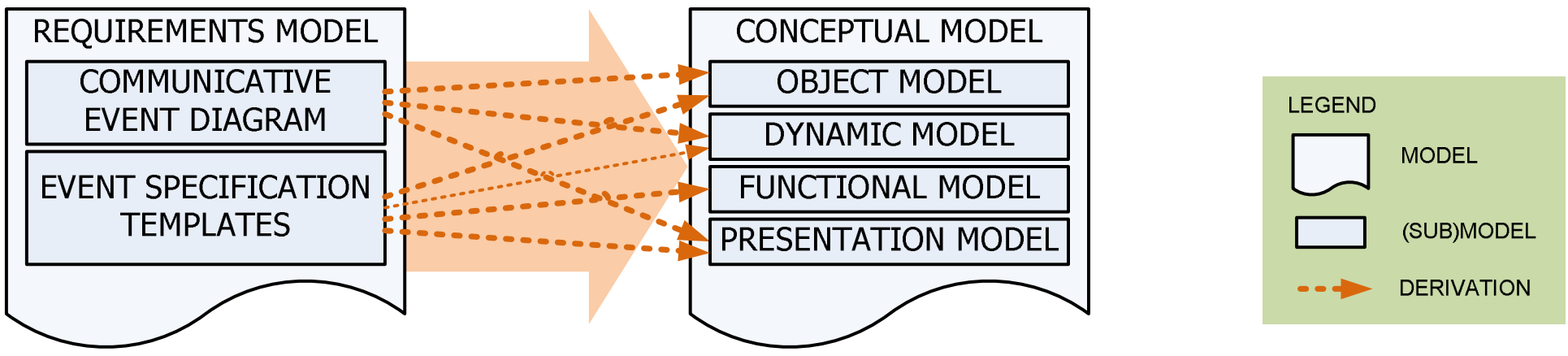

Figure 1. Derivation strategy from Communication Analysis to OO-Method models

The main step consists of processing the message structures in order to derive the class diagram view that corresponds to each communicative event. We refer as class diagram view to a portion of the class diagram that includes one or several portions of classes and structural relationships among them. By portion of a class we mean some or all of the attributes and services of the class (including their properties). The concept of class diagram view with regard to conceptual models is analogous to the concept of relational view with regard to relational database schemas [4].

Each communicative event derives a different class diagram view. The derived views are integrated to obtain the complete class diagram, the same way that different relational views are integrated to obtain a single database schema [4]. This integration can be approached in two ways [5]: post-process view integration or incremental view integration; we opt for the second. The procedure consists of the following steps. First, the communicative events are sorted. Then, the events are processed in order, one by one. The class diagram view that results from processing each communicative event is integrated into the class diagram under construction. Figure 2 exemplifies this approach. Note that one communicative event can affect one or several classes (event IV), and one class can be affected by one or many communicative events (class C).

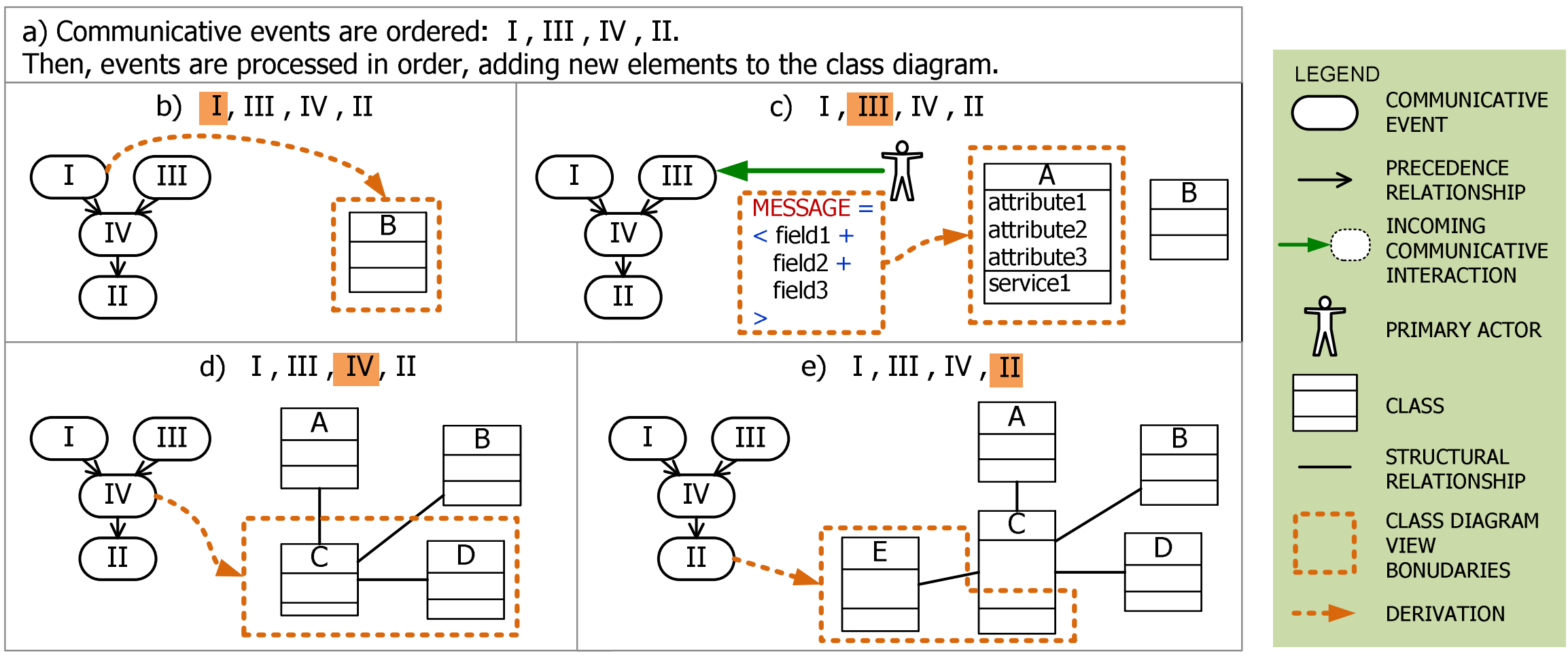

Figure 2. Simplified example of incremental class diagram view integration

By processing the communicative events in a predefined order and by performing incremental view integration, we intend to prevent a situation in which a newly-derived class diagram view refers to a class diagram element that has not yet been derived (e.g. trying to add an attribute to an inexistent class). In any case, if the requirements model suffers incompleteness, this situation cannot be fully prevented.

## 2.2. Running example

This section presents a simple case that is used below to illustrate the derivation rules. The illustrative case corresponds to part of the business processes of a fictional hospital named University Hospital Santiago Grisolía.

The University Hospital Santiago Grisolía is a state entity that provides medical services to its patients. The hospital mission is to improve the health of its patients by providing a personalised and sensitive healthcare, for which purpose many business processes have been carefully defined. One of them is the *Medical treatment* process. Most patients go to the hospital to ask an appointment with a doctor. Doctors prescribe medical treatments to treat the illnesses of the patients[1]. When a patient keeps an appointment with the doctor, the doctor specifies the medical treatment, by filling in the corresponding form. In the medical treatment form, the doctor specifies the initial and final date of the treatment, the nurse that will provide the medicaments, and the list of medicaments, specifying dosage, frequency, pain scale and sedation scale. Later on, the nurse assigns the dispensary where the medicaments will be provided, and she determines the delivery date (i.e. the date in which the patient can pick them up).

Figure 3 presents part of the communicative event diagram (CED) of the *Medical treatment* business process. Note that some communicative events that appear in the Figure 3 have precedent events that belong to other business processes (*App 1*, *Dis 1*, *Nur 1* and *Med 1* are events that belong to other business processes). These business processes are specified separately; i.e. they have their own CED and event specification templates {España, 2009 #1}.

[1] Obviously, in actual work practices, doctors make decisions according to the results of the anamnesis (i.e. the interview by which the doctor elicits information from the patient) and the appropriate medical tests (e.g. X-rays, blood tests, etc.). The reader should take into account that the illustrative case intends to be very simple, even if it ends up being simplistic. A detailed application of the derivation technique on a more realistic problem can be found in [1].

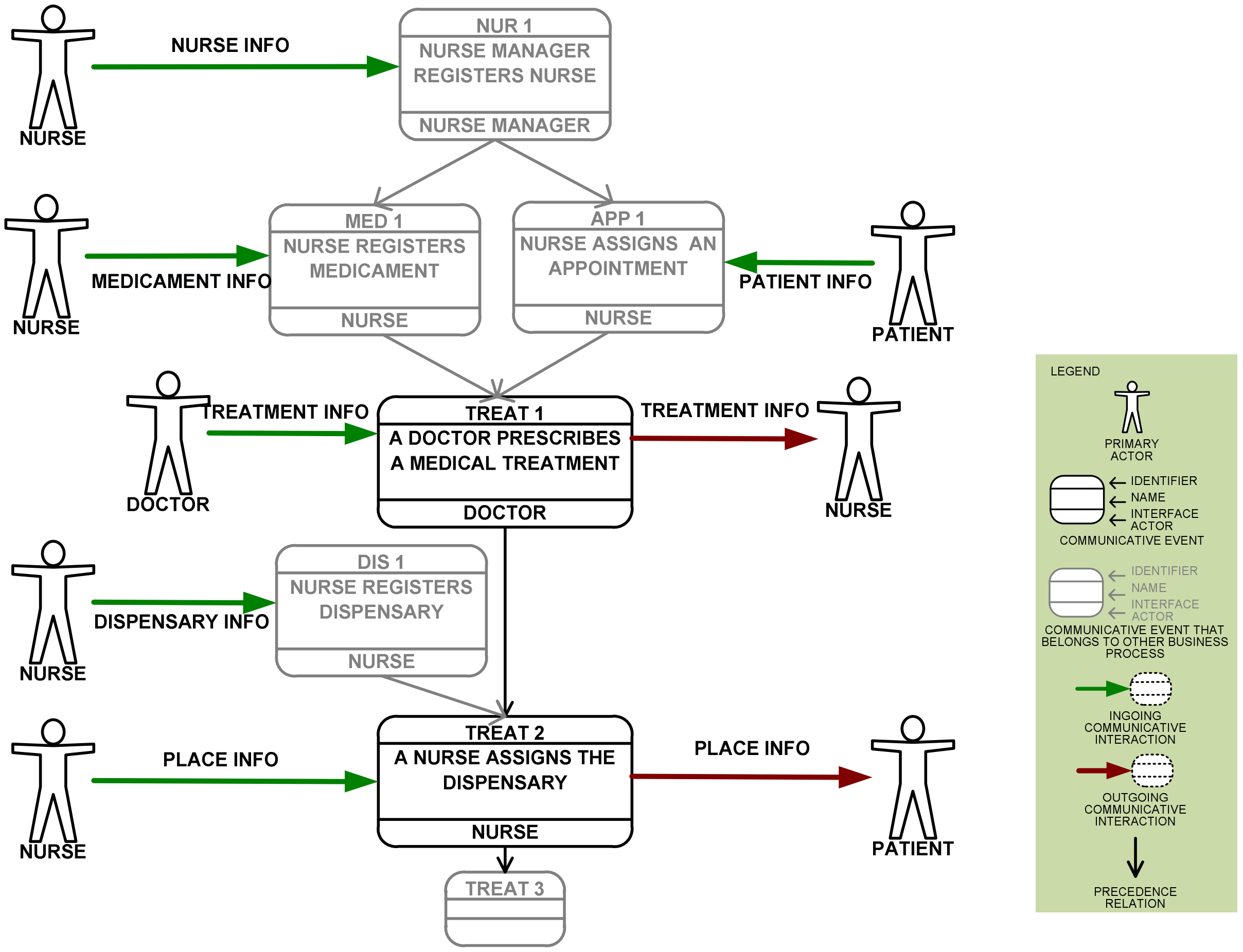

Figure 3. Comunivative event diagram of the Medical treatment business process (*TREAT*)

The event description templates for the communicative events *TREAT 1* and *TREAT 2* are included below.

***<u>TREAT 1. A doctor prescribes a medical treatment</u>***

## 1. General information

### Goals

The objective of the hospital is to attend the patients when they request medical services. From the point of view of the information system, the objective of this event is to record the medical treatment that is prescribed to a patient.

### Description

Most patients go to the University Hospital Santiago Grisolía to ask for an appointment with a doctor. Doctors prescribe medical treatments to treat the illnesses of the patients. When a patient keeps the appointment with the doctor, the doctor prescribes the medical treatment and fills out a medical treatment form. In the medical treatment form, the doctor specifies the initial and final date of the treatment, the nurse that will provide the medicaments, and the list of medicaments, specifying dosage, frequency, pain scale and sedation scale.

## 2. Contact requirements

### Actor responsabilities

- **Primary actor:** Doctor
- **Communication cannel:** In person
- **Interface actor:** Doctor

## 3. Communication requirements

### Message structure

Table 1. Message structure of the communicative event *TREAT 1*

| FIELD | OP | DOMAIN | EXAMPLE VALUE |
|---|---|---|---|
| MEDICAL TREATMENT = | | | |
| < Treatment number + | g | number | 26411 |
| Initial date + | i | date | 01-08-2004 |
| Final date + | i | date | 01-08-2005 |
| Patient + | i | Patient | 842133-W, Richard Pain |
| Nurse + | i | Nurse | APCB |
| Comments + | | Text | |
| MEDICATIONS = | | | |
| { MEDICATION = | | | |
| < Medicament + | | Medicament | Folic Acid, Tab 1Mg |
| Dosage + | i | text | 1 tab 1Mg |
| Frequency + | i | text | Every morning |
| Pain scale + | i | text | No pain |
| Sedation scale | i | text | Wide awake |
| > | | | |
| } | | | |
| > | | | |

**Structural restrictions**
One medical treatment is specific for exactly one patient.

One patient can have many medical treatments.

**Contextual restrictions**
Medical treatments are identified by the treatment number.

### 4. Reaction requirements

**Treatments**
The medical treatment is recorded.

**Linked communications**
The nurse is informed of the medical treatment has been prescribed.

## ***TREAT 2. A nurse assigns the dispensary***

### 1. General information

**Goals**
The objective of the hospital is to assign the most appropriate dispensary where the medicaments will be provided.

**Description**
The nurse assigns the dispensary where the treatment will be provided, and she assigns the treatment delivery date.

### 2. Contact requirements

**Actor responsabilities**

- **Primary actor:** Nurse
- **Communication cannel:** In person
- **Interface actor:** Nurse

### 3. Communication requirements

**Message structure**

Table 2. Message structure of the communicative event *TREAT 2*

| FIELD | OP | DOMAIN | EXAMPLE VALUE |
|---|---|---|---|
| DISPENSARY = | | | |
| < Treatment + | i | Medical treatment | 26411 |
| Delivery date + | i | date | 03-08-2004 |
| Dispensary + | i | Dispensary | SG Lab |
| > | | | |

**Structural restrictions**

One dispensary can be assigned to many medical treatments.

One medical treatment is assigned to exactly one dispensary.

**4. Reaction requirements**

**Treatments**

The medical treatment is recorded.

**Linked communications**

The patient is informed of dispensary and delivery date.

## 2.3. Derivation guidelines

In the following, we provide the derivation guidelines to construct the class diagram.

First, the communicative event diagram is extended with precedent communicative events so as to obtain a rooted directed graph (see rule **OM1**)[2].

| **Rule OM1**. Extension of the communicative event diagram | |
|---|---|
| **Preconditions** | |
| | None |
| **Textual explanation** | |
| 1 | Include any communicative event that is not included in the communicative event diagram being processed, but it precedes a communicative event which is included in the diagram (review other precedent business processes). |
| 2 | Repeat step 1 until no more events are included. |
| **Example** | |
| Figure 4 presents an example of an extension process of a communicative event diagram. Business process *A* will be extended with the communicative events that are needed from other related business processes in order to obtain a rooted directed graph. | |

[2] To illustrate this derivation rule, several communicative event diagrams are needed; in order to avoid complicating the running example, we illustrate the application of this rule using a different example.

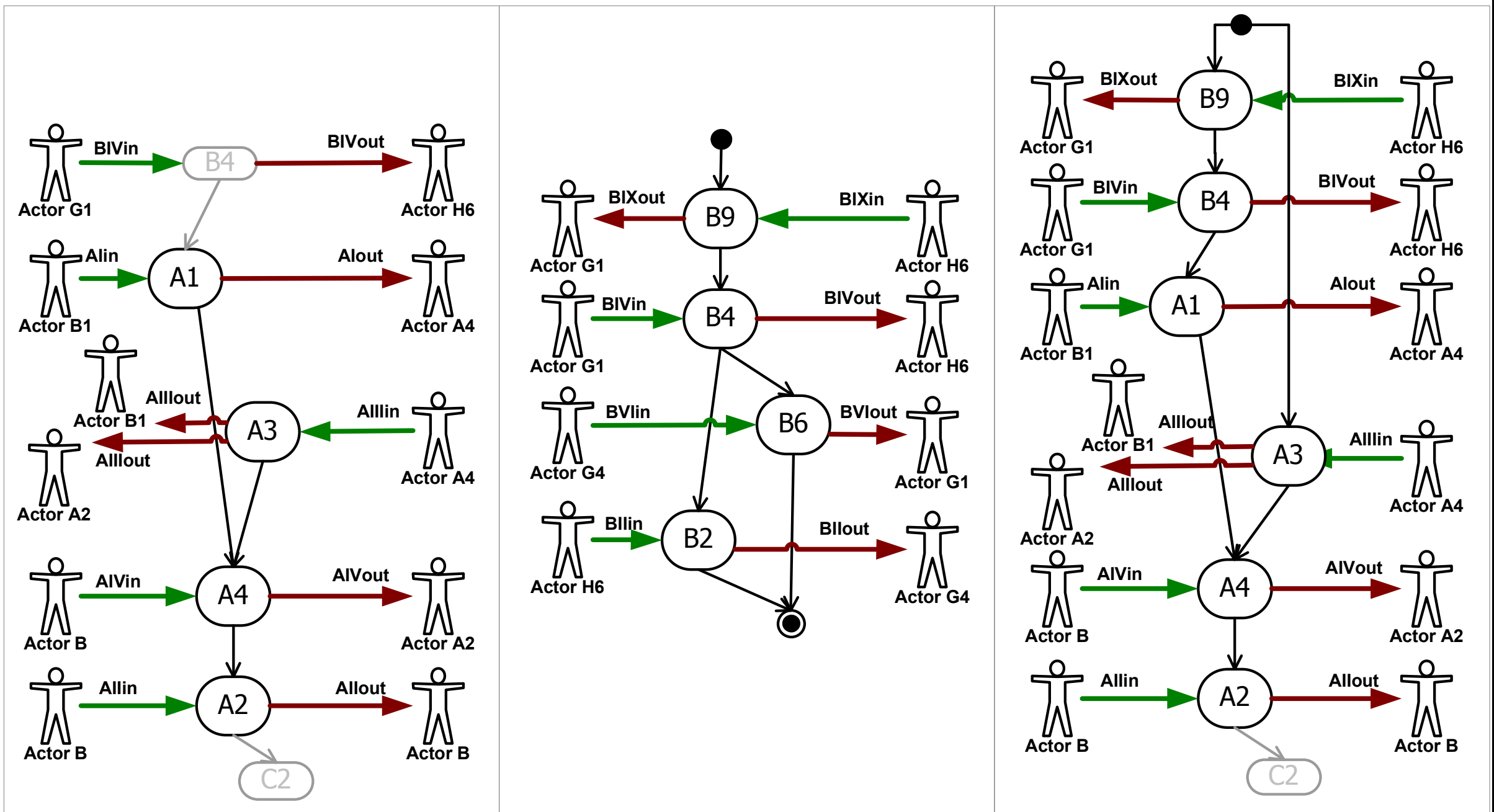

a) Communicative event diagram for *A* business process

b) Communicative event diagram for B business process

c) Extended communicative event diagram extended for *A* business process with B business process

Figure 4. Simplified example of business process extension

It is also noteworthy that the communicative event diagram of business process *A* (see Figure 4.a) does not include a start node, while B does (See Figure 4.b). A communicative event is preceded by the start node to remark that it is not preceded by any communicative event. Each of the communicative events of business process *A* has at least one precedent event, either from business process itself or from a different business process (e.g. *A4* has *A1* and *A3* as precedents; *A1* has the communicative event *B4* as precedent; in the *B* business process B4 is preceded by *B9*).

A similar line of reasoning is done for the end node. A communicative event is followed by an end node to remark that it does not precede any other communicative event. In business process *A*, each communicative event precedes another event (event *A2* precedes *C2*, although it is out of the scope of this illustrative example. In contrast, in the business process *B*, *B2* and *B6* do not precede any other communicative event; therefore, *B2* and *B6* are followed by an end node.

Note that the communicative event *C2* is out of the scope of this analysis, and in the moment of extend the communicative event diagram, this communicative event is not processed.

Figure 4.c presents the extended communicative event diagram for business process *A* (this includes the communicative events that are precedents to any of the events in business process *A*).The start node and end node can be omitted of the sake of simplicity (i.e. to avoid many line crossings).

Figure 5 presents an extended communicative event diagram of business process *A* that omits the start node

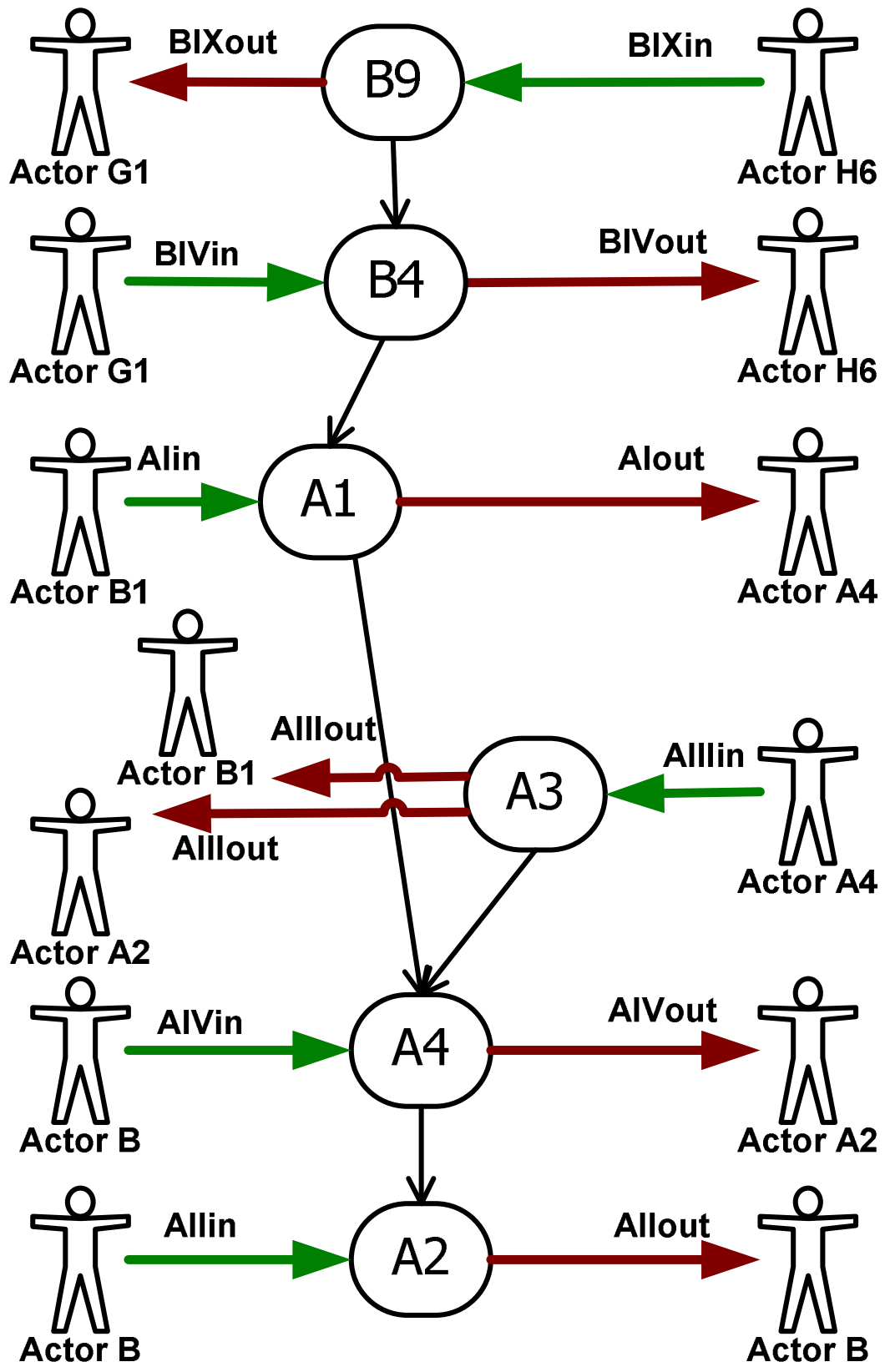

Figure 5. Extended diagram that omits the start node for the sake of clarity

Now returning to the running example, in the business process *Medical treatment*, as a pre-processing step of the message structures, the analyst marks those reference fields which indicate that (part of) a business object is being extended with new information. These marks can be supported by means of a new Boolean property for reference fields named *Extends business object* in the message structure description (see rule **OM2**).

| **Rule OM2**. Marking of reference fields | |
|---|---|
| **Preconditions** | |
| | Rule **OM1** has already been applied. |
| **Textual explanation** | |
| 1 | Add a new Boolean property (i.e. column) named *Extends business object* to every message structure of the requirements model. The new property only applies to reference fields. |
| 2 | Mark each reference field that indicates that an aggregation substructure is describing the provision of new information about an already-known business object. The mark consists of setting the new property to True. |
| **Example** | |
| When this rule is applied to the communicative event *TREAT 1. A doctor prescribes a medical treatment*, no changes are made, because the event does not extend a business process but actually creates a new one. Therefore, we focus on communicative event *TREAT 2. A nurse assigns the dispensary*, please see the event description template on page 5. Given to the meaning of the communicative event (its pragmatic consequences), the reference field Treatment actually specifies that the nurse selects an existing medical treatment form in order to make a decision. She decides when and where the treatment | |

will be delivered. Then she adds this information to the form.

Table 3. Message structure of the communicative event *TREAT 2* marked according to rule **OM2**

| FIELD | OP | DOMAIN | EXAMPLE VALUE | EXTENDS BUSINESS OBJECT |
|---|---|---|---|---|
| DISPENSARY =<br>< Treatment +<br>Delivery date +<br>Dispensary +<br>> | <br>i<br>i<br>i | <br>Medical treatment<br>date<br>Dispensary | <br>26411<br>03-08-2004<br>SG Lab | <br>True |

Therefore, the reference field Treatment is referencing a business object being extended with new information; namely, MEDICAL_TREATMENT. According to this derivation rule, Treatment should have its property *Extends business object* set to True.

**Note**

Only one reference field per aggregation substructure can be marked.

It is not necessary to mark as False the remaining reference fields (False is considered the default value).

Then, the communicative events depicted in the extended communicative event diagram are ordered according to the precedence relationships (see Rule **OM3**). For this purpose, a partially ordered set of communicative events is obtained by removing the loopbacks that might appear in the extended communicative event diagram. The remaining precedence relations define a strict partial order among the events. Any known procedure for topological sorting can be used for obtaining the ordered list of events; for instance, the algorithm proposed by Kahn [6]. Often many solutions (a.k.a. linear extensions) are possible; any solution is suitable for derivation purposes and the choice is not expected to influence efficiency.

**Rule OM3**. Event ordering

**Preconditions**

Rule **OM1** has already been applied.

**Textual explanation**

| | |
|---|---|
| 1 | Remove the loopbacks that might appear in the extended communicative event diagram and remove every other model element except for precedence relationships and communicative events. |
| 2 | *L* = empty list that will contain a sorted list of communicative events. |
| 3 | *S* = set of communicative events whose incoming precedence relationship comes from the start node. |
| 4 | While *S* is not empty |
| 5 | Remove a communicative event *e* from *S*. |
| 6 | Insert *e* into *L*. |
| 7 | For each communicative event *f* with a precedence relationship *p* from *e* to *f* do |
| 8 | Remove precedence relationship *p* from the diagram. |
| 9 | If *f* has no other incoming precedence relationships then insert *f* into *S*. |

**Example**

Figure 6 a shows the poset[3] of communicative events of a given diagram[4]. Figure 6.b presents all the ordered lists of events that are compatible with the partial order defined by the precedence relations. Take into account that any of these solutions is suitable and, thus, only one of the lists needs to be obtained.

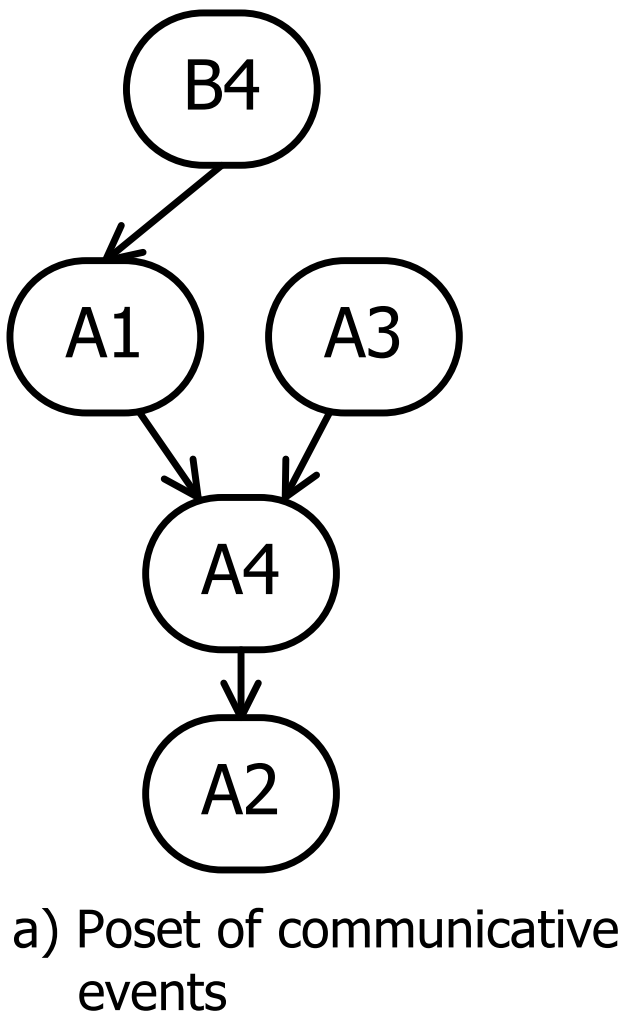

B4, A1, A3, A4, A2
B4, A3, A1, A4, A2
A3, B4, A1, A4, A2

b) Suitable ordered lists of events

Figure 6. Example of event ordering

Note that the list A1, B4, A3, A4, A3 would not be suitable because A1 cannot appear before B4 (there is a precedence relation between them in the opposite direction).

After applying the rule OM3, we have a suitable ordered list. This list guides the order for processing the communicative events. For each event, its message structure is processed to obtain the class diagram view that corresponds to the communicative event. In the running example of the hospital, we do not present the whole derivation; however, a possible order in which the communicative events in Figure 3 can be processed is: *NUR 1*, *MED 1*, *APP 1*, *TREAT 1*, *DIS 1*, *TREAT 2*.

The message structure presented in Table 1 consists of an initial aggregation structure named MEDICAL TREATMENT that includes an iteration substructure named MEDICATIONS. This iteration substructure has an aggregation substructure inside of it (MEDICATION).

Each aggregation substructure that does not contain a reference field that has been marked according to **OM2**, leads to the derivation of a new class (see rule **OM4** ).

[3] A poset consists of a set together with a binary relation that indicates that, for certain pairs of elements in the set, one of the elements precedes the other. In our case, the binary relation is defined by the precedence relationships. If a communicative event A precedes another communicative event B then we can consider that A<B. The start node (even if it is implicit) is the *least element* of the poset.

[4] Again, a simpler example than the hospital case is used, for the sake of brevity.

| **Rule OM4**. Class creation | |
|---|---|
| **Preconditions** | |
| | Rule **OM2** has already been applied. |
| **Textual explanation** | |
| 1 | For each aggregation substructure that *does not* contain a reference field whose property *Extends business object* is True (i.e. the field has been marked according to **OM2**), create a new class. |
| **Example** | |
| The two aggregation substructures showed in Table 1 lead to the derivation of the classes MEDICAL_TREATMENT and MEDICATION respectively. | |

The name of the aggregation substructure can be used to name the class (see rule **OM5**), spaces are replaced by underscores; however, the analyst can decide to give the class a different name.

| **Rule OM5**. Assign the class a name | |
|---|---|
| **Preconditions** | |
| | Rule **OM4** has already been applied. |
| **Textual explanation** | |
| 1 | For each newly-derived class, assign to the name of the class the name of its corresponding aggregation substructure. |
| 2 | Replace spaces by underscores and use uppercase letters. |
| 3 | In any case, the analyst can choose to give a different name to a class. |
| **Example** | |
| Figure 7 shows the derivation of classes from aggregation substructures (see rule **OM4**) and assignation of a name to each class. | |

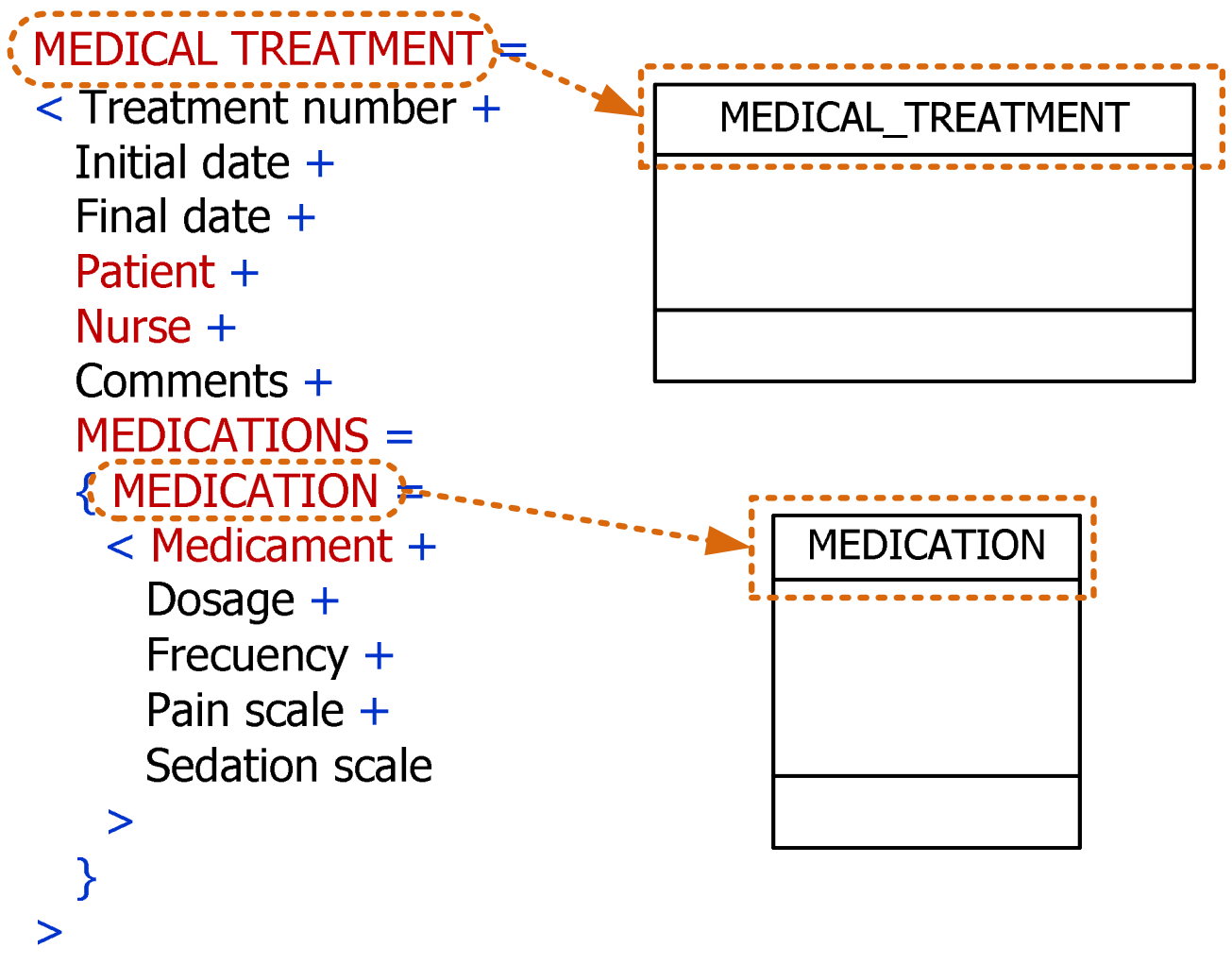

Figure 7. Creation of classes and name assignation

The aggregation substructures MEDICAL TREATMENT and MEDICATION include data fields. For instance, the complex substructure MEDICAL TREATMENT includes the data fields Treatment number, Initial date Final date and Comments. For each data field of a substructure, its corresponding class is added an attribute (see rule **OM6**).

| **Rule OM6**. Add attributes to each class | |
|---|---|
| **Preconditions** | |
| | Rule **OM5** has already been applied. |
| **Textual explanation** | |
| 1 | For each data field of the substructure, add an attribute to the previously-derived class. |
| **Example** | |
| The data field Treatment number leads to adding the attribute treatment_number to the class MEDICAL_TREATMENT; the same is done for the data fields Initial date, Final date and Comments. | |

The names of the attributes are derived from the names of the data fields (see rule **OM7**), spaces are replaced by underscores and lowercase letters are used.

| **Rule OM7**. Assign a name to the attributes | |
|---|---|
| **Preconditions** | |
| | Rule **OM6** has already been applied. |
| **Textual explanation** | |
| 1 | Assign the name to attribute according to its corresponding data field (i.e. the data field from which the attribute is derived). |
| 2 | Replace spaces by underscores and use lowercase letters. |
| 3 | In any case, the analyst can choose to give a different name to an attribute. |
| **Example** | |
| The Treatment number data field leads to adding the attribute treatment_number to the class MEDICAL_TREATMENT; the same are done for the data fields Initial date Final date and Comments (see Figure 8). | |

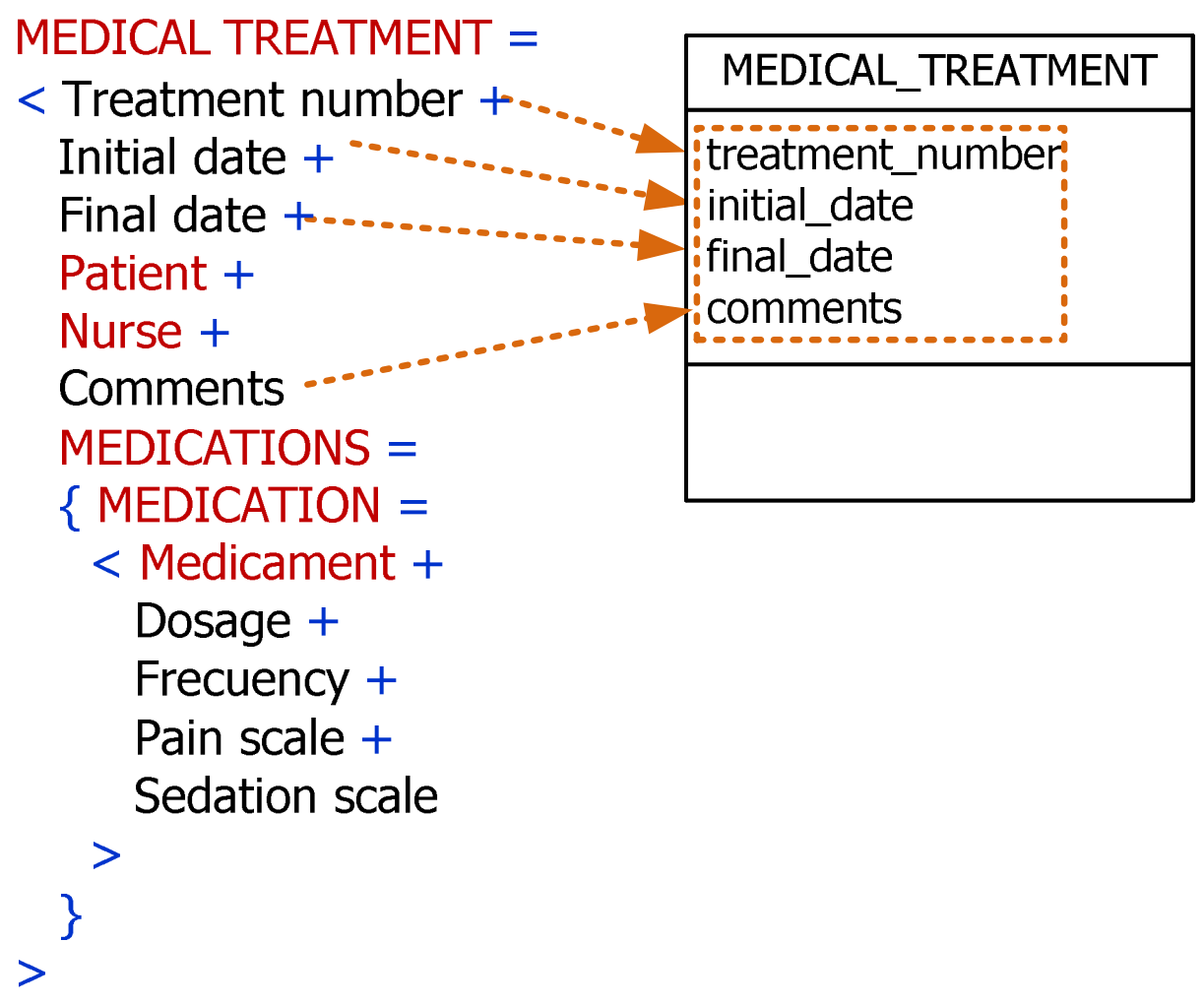

Figure 8. Creation of the attributes treatment_number, initial_date, final_date and Comments

The information contained in the event specification templates (e.g. message structures, structural restrictions and contextual restrictions) is used to derive the class identifier and to derive other properties such as the attribute data types, their sizes, whether they allow null values, etc.

Table 4 presents the specification of the attributes of the class MEDICAL_TREATMENT after processing the event *TREAT 1*. In the following, we explain how this information is derived.

Table 4. Specification of the attributes of the class MEDICAL_TREATMENT after processing the event *TREAT 1*[5]

| Attribute name | Id | Attribute type | Data type | Size | Requested | Null allowed |
|---|---|---|---|---|---|---|
| treatment_number | yes | Constant | Autonumeric | - | yes | no |
| initial_date | no | variable | Date | - | yes | no |
| final_date | no | Variable | Date | - | yes | no |
| comments | no | Variable | String | 200 | yes | yes |

The requirements model should specify how the organisational actors identify business objects of each kind. If available, this information is specified as a contextual restriction and it offers guidance for selecting which attributes constitute the class identifier (see **OM8**). In the absence of such restrictions, then the analyst should ask the users or decide relying on his/her own judgement.

| **Rule OM8**. Select class identifier | |
|---|---|
| **Preconditions** | |
| | Rule **OM7** has already been applied. |
| **Textual explanation** | |
| 1 | In case there exists a contextual restriction in the event specification template that indicates which message structure field (or set of fields) should be used to identify a given business object (or a part of it), then this information is used to select which attributes constitute the class identifier. |
| 2 | In the absence of such restriction, the analyst should ask the users or decide relying on his/her own judgement. An option is to create an artificial autonumeric identifier and give it the name class_name_id (i.e., the class name with the suffix _id). |
| **Example** | |
| According to the event specification template (See page 4), the class MEDICAL_TREATMENT is identified by a Treatment number, so the attribute treatment_number is designated the class identifier. In Table 4 is specified in the column **Id**. | |

The attribute type should be asked to the users taking into account that, once a constant attribute has been initialised, its value cannot be modified. In contrast, variable attributes can be modified (as long as the proper class services are defined)[6] [7]. However, by default, all attributes that are part of the class identifier are defined as constant; the rest are defined as variable since it is the less restrictive option (see **OM9**).

[5] These tables are not part of the requirements model; they are part of the conceptual model. At the beginning of the derivation process, such tables do not exist. Whenever a class is created (by means of applying rule **OM4**), an empty table is created. Later, whenever a class attribute is derived (by means of applying rule **OM6**), an empty row in the corresponding table is created. As the properties of the attributes are derived, the columns are filled.

| **Rule OM9**. Specify attribute type | |
|---|---|
| **Preconditions** | |
| | Rule **OM7** has already been applied |
| **Textual explanation** | |
| 1 | By default, the attributes that are part of the class identifier are defined as constant, the rest are defined as variables. |
| 2 | However, the analyst should ask the users the following question |
| 3 | "Consider that this piece of information about a business object (indicate the attribute name) is set a value in a given moment; is it possible that this information needs to be changed later?" |
| 4 | If the response is "Yes, it should be possible to change this piece information later" then the attribute type is variable. |
| 5 | If the response is "No, the initial value is never changed" then the attribute type is constant. |
| **Example** | |
| The attribute treatment_number is part of the class identifier. This means that the attribute is of type Constant. The information about the attribute type is specified in Table 4. The remaning attributes are of type variable by default. In any case, the analyst should ask the users whether the information related to these attributes needs to be changed later or not. | |

The data types of the attributes are derived from the domains of their corresponding message structure data fields (see rule **OM10**). Communication Analysis prescribes a few basic domains for data fields (which serve the purpose of clarifying the meaning of the messages), whereas the OO-Method offers a wider selection of data types for attributes (code generation is intended).

Table 5 defines a correspondence between Communication Analysis data field domains and OO-Method attribute data types, thus offering some guidance for the conversion.

Table 5. Conversion table for attribute data types

| **Communication Analysis Data field domain** | **OO-Method Attribute data type**[7] |
|---|---|
| number | Nat, Int, Real, Autonumeric[8] |

[6] Take into account that, according to the *OLIVA**NOVA***, constant attributes can only be initialised by means of the creation service; therefore, an attribute not being requested upon creation and of type constant is useless. This is a restriction of the tool and it does not apply to the OO-Method. According to the method, a constant attribute can have its value initialised in an edition service as well.

[7] We clarify the meaning of some data types that are not straightforward: *Nat* stands for natural number, *Int* for integer, *Autonumeric* refers to a sequence of natural numbers that is incremented automatically every time an instance is created, *String* refers to an array of characters of a specified length, *Text* refers to a multi-line text, *Bool* stands for Boolean, *Blob* stands for binary large object.

[8] Take into account that the data type Autonumeric should only be used for constant attributes (e.g. class identifiers).

| | |
|---|---|
| text | String, Text |
| date, time | Time, Date, DateTime |
| money | Real |
| *not considered* | Bool, Image, Blob |

| **Rule OM10**. Specify attribute data type | |
|---|---|
| **Preconditions** | |
| | Rule **OM7** has already been applied |
| **Textual explanation** | |
| 1 | Assign the attribute data type according to the conversion table for attribute data types (see Table 5). |
| 2 | Some data field domains can be converted to several attribute data types; in these cases, the analysts should apply their criteria. However, the data types that are underlined are the ones that are recommended in case of doubt. |
| 3 | In case a String data type is chosen, then the Size (the maximum amount of characters that the string attribute will allow to store) needs to be defined. For this purpose, the analyst asks the users or decides relying on his/her own judgement. |
| **Example** | |
| In order to derive the data type of the attribute treatment_number of class MEDICAL_TREATMENT, we first trace back the message structure data field from where the attribute has been derived: it is Treatment number (see Figure 8). The domain of Treatment number is Number (see page 5). This domain is looked up in the first column of Table 5 (first row) and one of the corresponding OO-method attribute data types can be chosen (either Nat, Int, Real or Autonumeric). Autonumeric is chosen because, in this case, an automatic sequence of natural numbers is convenient for identifying medical treatment forms. Finally, this property of the attribute treatment_number is specified in Table 4. Column **Data type**.<br><br>Similarly, the data fields that correspond to the attributes initial_date and final_date are Initial date and Final date, respectively. The domain of both data fields is date, so a Date time data type is chosen for the attributes. With regards to the attribute comments, it has been derived from the data field Comments, which has a text domain; therefore a data type String is chosen. Since this data type needs a size, the analyst asked the doctors how long were the longest comments they usually write in medical treatment forms: an appropriate size is 200 characters. The size is specified in Table 4. Column **Size**. | |

In a newly-derived class, all the attributes are set as requested at creation time (see rule **OM11**).

| **Rule OM11**. Set attributes as requested at creation time | |
|---|---|
| **Preconditions** | |
| | Rule **OM7** has already been applied. |
| **Textual explanation** | |
| 1 | In case the class has been created during the processing of the current communicative event (i.e. it is a newly-derived class), then set every attribute as *Requested upon creation*. |
| 2 | On the contrary, the new attributes that are added to a class when it is extended (i.e. when processing a successor event that affects an already-existing business object) should not be requested upon creation, because the information they contain is not provided when the instance is created, but in a later moment. |
| **Example** | |
| Since the class MEDICAL_TREATMENT is derived for the first time during the processing of the event *TREAT 1*, therefore all of the newly-derived attributes are set as *Requested upon creation*. This has been indicated in Table 4, column **Requested**. | |

Whether an attribute should allow null values or not needs to be asked to the users (or the analyst should decide relying on his/her own judgement). However, by default, all attributes that are part of the class identifier are added a restriction so that they do not allow null values; the rest of the attributes are set to allow null values (see rule **OM12**).

| **Rule OM12**. Allow null values for the attributes | |
|---|---|
| **Preconditions** | |
| | Rule **OM7** has already been applied. |
| **Textual explanation** | |
| 1 | If the attribute is part of the class identifier, then set the value of the property **Null allowed** to "No". |
| 2 | Else, set the value to "Yes". |
| **Example** | |
| Since the class identifier is treatment_number, then this attribute is set to not allow null values. With regards to initial_date and final_date, although they are not part of the class identifier, the analyst decides to make them compulsory based on the comments by the users. The attribute comments is set to allow null values, because it is not mandatory that the doctor comments on the medical treatment. | |

The nesting of complex substructures leads to the derivation of structural relationships between the classes that correspond to the substructures (rule **OM13**).

| **Rule OM13**. Nesting substructures derives structural relationships | |
|---|---|
| **Preconditions** | |
| | Rule **OM5** has already applied. |
| **Textual explanation** | |
| 1 | Create a structural relationship between the classes that correspond to every pair of nested aggregation substructures. |
| **Example** | |
| The aggregation substructure MEDICATIONS is part of the aggregation substructure MEDICAL_TREATMENT, then, the corresponding classes | |

(MEDICAL_TREATMENT and MEDICATION, respectively) are related by means of a structural relationship (see Figure 9).

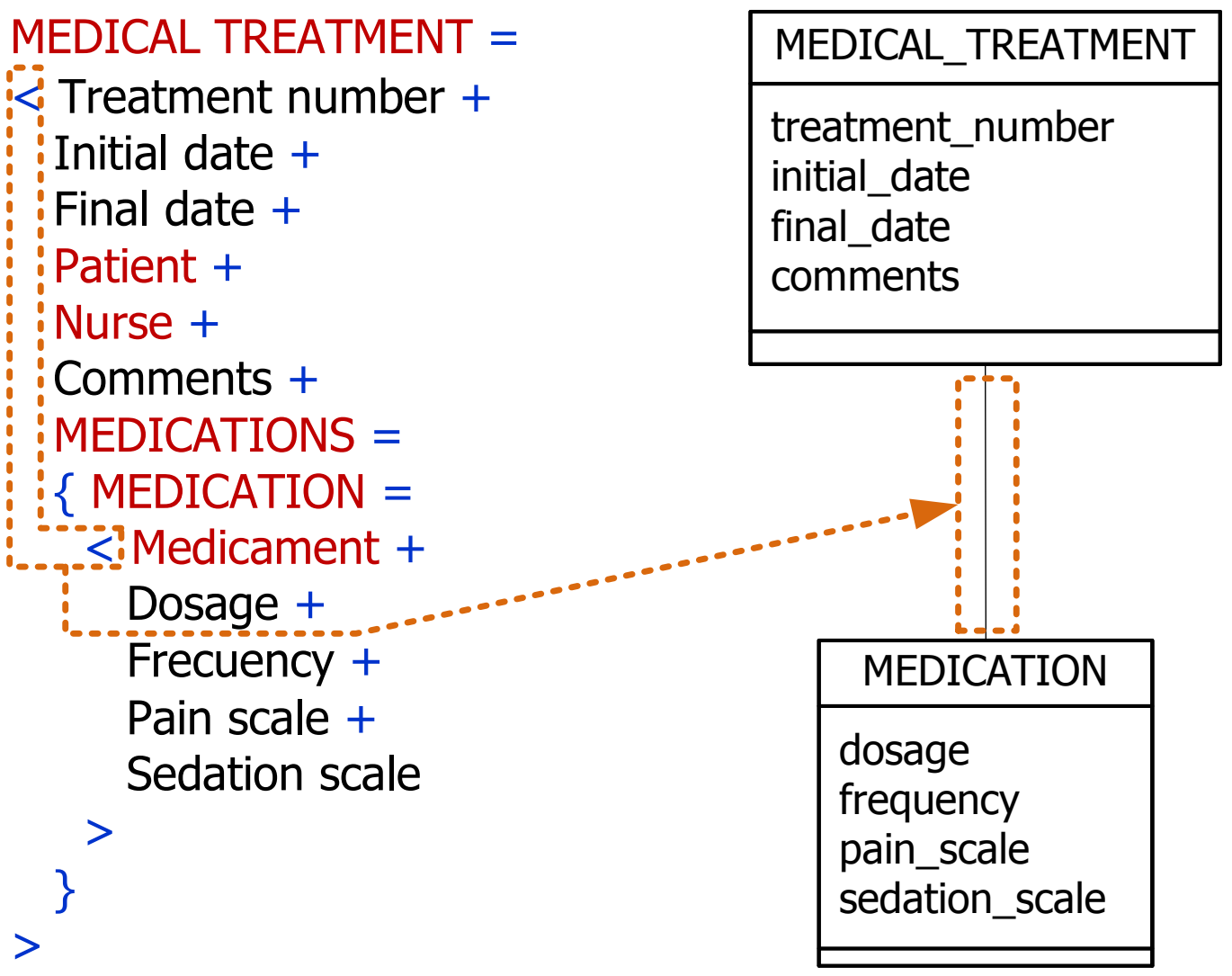

Figure 9. Relationship between the classes MEDICAL_TREATMENT and MEDICATION

For the structural relationships derived from substructure nesting, it is possible to also derive the cardinality[9] (see rule **OM14**).

| **Rule OM14**. Assign cardinality (analysis of aggregation substructures) | |
|---|---|
| **Preconditions** | |
| | Rule **OM13** has already applied. |
| **Textual explanation** | |
| 1 | If the nested substructure is an aggregation substructure then specify that the maximum cardinality on the side of the referenced class is 1. |
| 2 | If it there is an iteration substructure in between, then this cardinality in M (many). |
| **Example** | |
| Since MEDICATIONS (the substructure that relates MEDICAL TREATMENT and MEDICATION) is an iteration substructure, then the structural relationship has maximum cardinality M on the side of class MEDICATION (i.e. the iteration specifies that one medical treatment can consist of the prescription of many medications); see Figure 10. | |

[9] We acknowledge that cardinality is often referred to as multiplicity since UML became a de-facto standard; however, OO-Method still uses this term [Pastor and Molina 2007]. Also, the notation in *OLIVA**NOVA*** departs from the UML: CLASSA minA:maxA --- minB:maxB CLASSB.

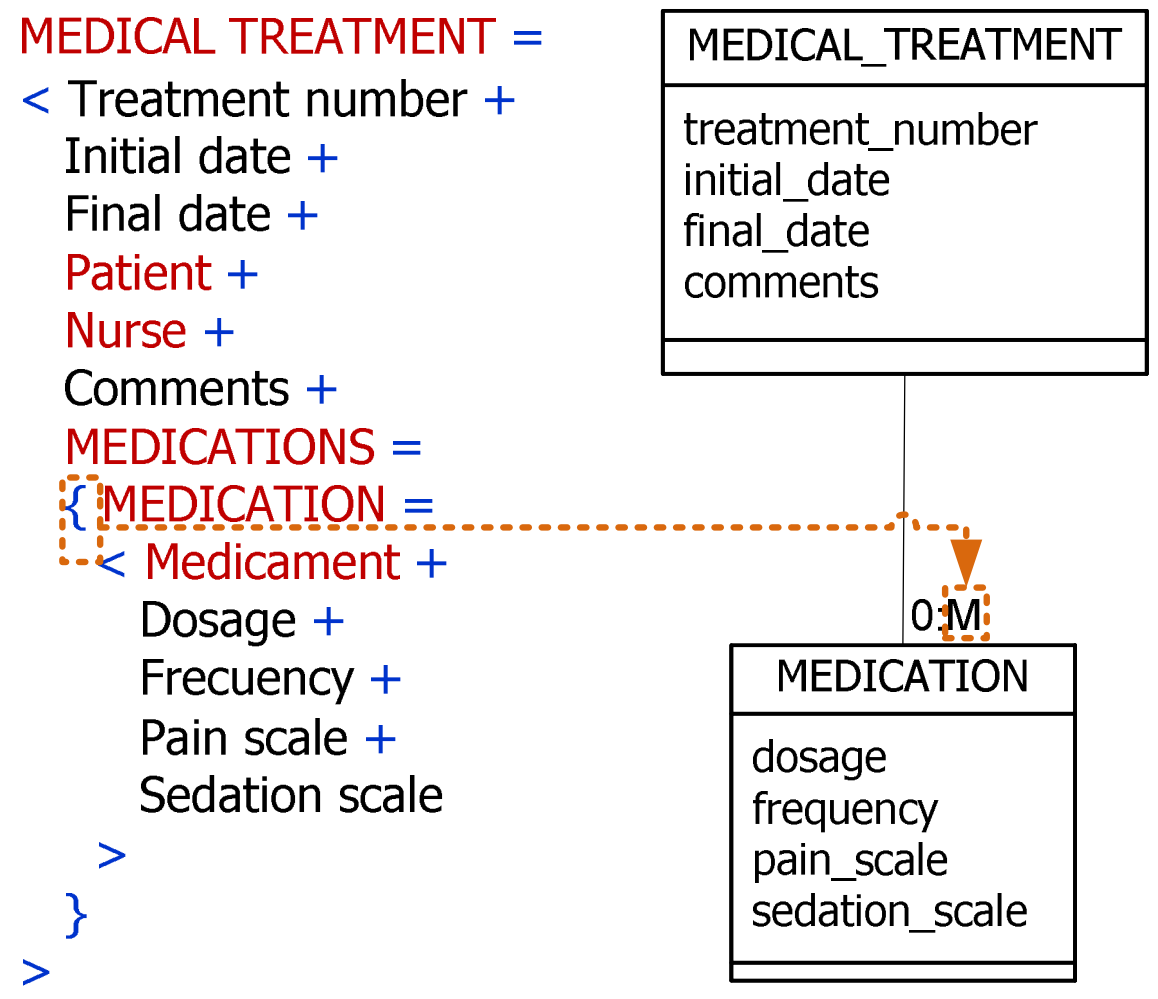

Figure 10. Addition of cardinalities to the relationship between classes MEDICAL_TREATMENT and MEDICATION

Now, the remaining cardinalities depend on structural restrictions that appear in the event description templates (see rule **OM15**).

| **Rule OM15**. Assign cardinality (analysis of structural restrictions) | |
|---|---|
| **Preconditions** | |
| | Rule **OM14** has already applied. |
| **Textual explanation** | |
| 1 | Assign the rest of the cardinalities according to structural restrictions included in the event description template. |
| 2 | In the absence of such restrictions, the analyst should ask the users or decide relying on his/her own judgement. |
| **Example** | |
| There is no structural restriction regarding the remaining cardinalities. Based on the analyst's experience, the cardinality on the side of MEDICAL_TREATMENT is 1:1 (see Figure 11).<br>Note that a possible structural restriction could have been "One medication prescription corresponds to one and only one medical treatment". This restriction would have led to the same cardinalities. | |

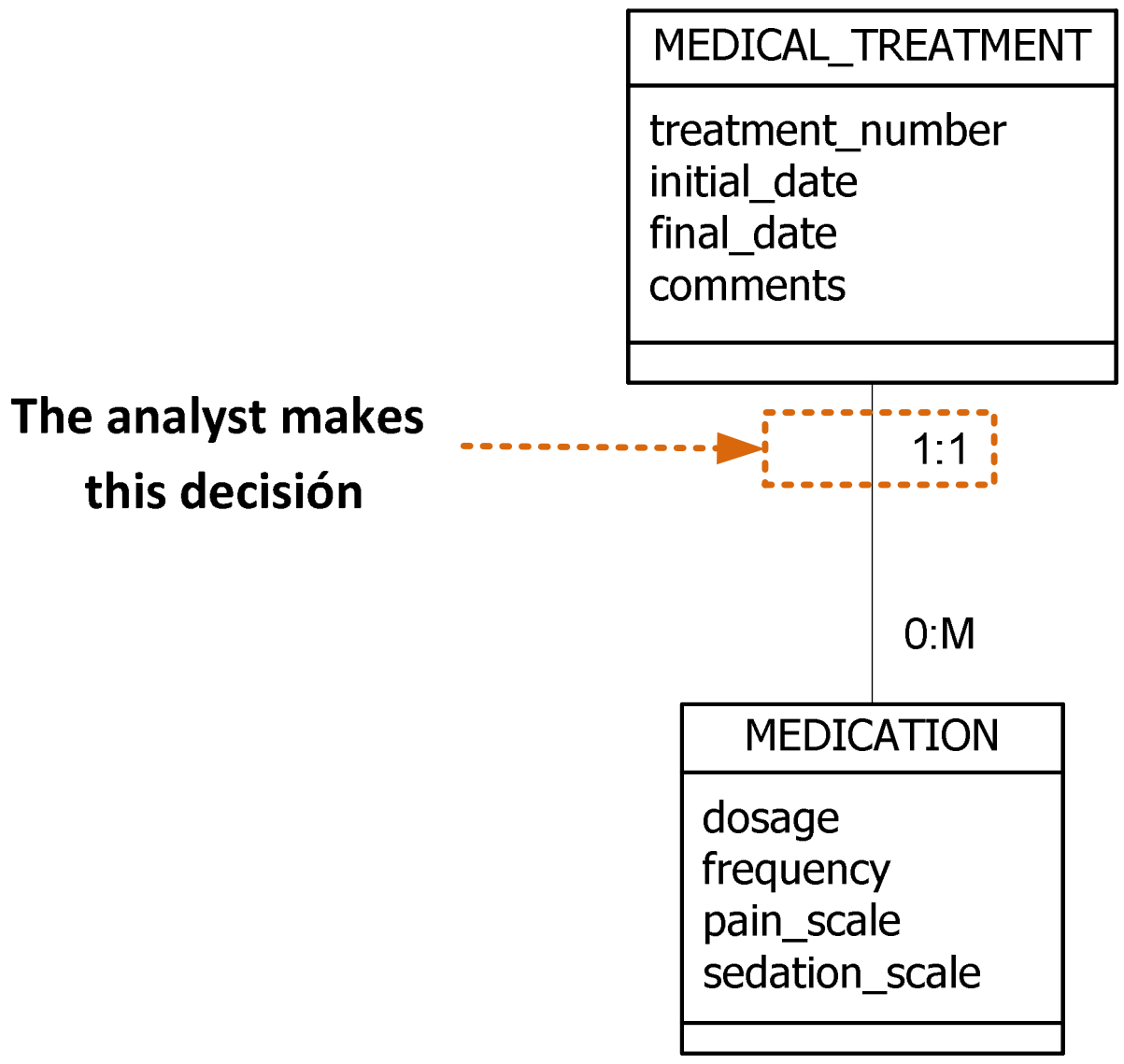

Figure 11. Addition of cardinalities for the class MEDICAL_TREATMENT

Reference fields lead to the derivation of structural relationships between the class that corresponds to the complex substructure containing the reference field and the class that corresponds to the reference field (see **OM16**).

| **Rule OM16.** Reference field derives structural relationship | |
|---|---|
| **Preconditions** | |
| | Rules **OM5** Has already applied. |
| **Textual explanation** | |
| 1 | For each reference field create a structural relationship between the class that correspond to the complex substructure containing the reference field and the class that corresponds to the reference field. |
| 2 | Note that, as long as the communicative events have been properly sorted and processed in order (see **OM3**), the classes that corresponds to the reference field have already been derived during a previous event processing. Otherwise, the requirements model suffers from incompleteness (e.g. a communicative event is missing). |
| **Example** | |
| At this moment of the derivation, the processing of previous communicative events has already led to the derivation of several classes (we depict them in grey in Figure 12).<br>The reference field Patient belongs to the complex substructure MEDICAL TREATMENT, whose corresponding class is MEDICAL_TREATMENT. The domain of Patient is Patient (see page 5); during the processing of communicative event *PAT 1*, the class PATIENT was derived. Therefore, a structural relationship is defined between the classes MEDICAL_TREATMENT and PATIENT.<br>A similar reasoning is done for the reference fields Nurse and Medicament. | |

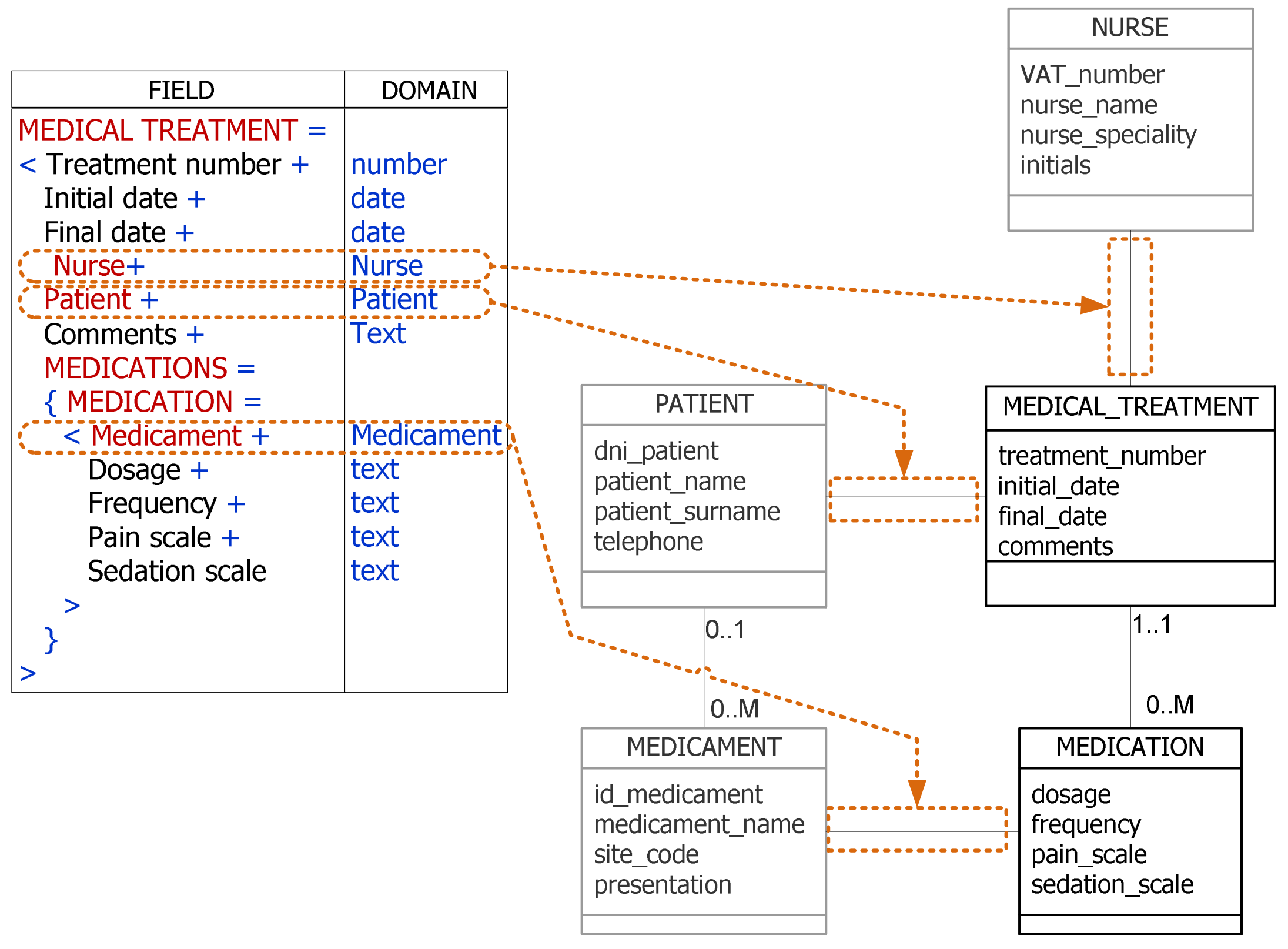

Figure 12. Creation of structural relationships derived from reference field

Then, the cardinalities for these structural relationships are assigned according to **OM17**

| **Rule OM17**. Assign cardinalities (analysis of reference fields) | |
|---|---|
| **Preconditions** | |
| | Rule **OM16** has already applied. |
| **Textual explanation** | |
| 1 | Assign the maximum cardinality 1 on the role side of the referenced class. Note that, if a higher cardinality had been intended, an iteration substructure should have been used instead of a reference field. |
| 2 | With regards to the minimum cardinality and the cardinalities on the other role side of the relationship, assign them according to the structural restrictions included in the event description template (see rule **OM15**). |
| **Example** | |
| Following the example, the message structure of *TREAT 1* includes a reference field that refers to a patient; it represents the patient for whom the medical treatment is prescribed. Thus, the maximum cardinality on the side of class PATIENT is 1. Also, there is a structural relationship in the event specification template stating that “A medical treatment is specific for exactly one patient” (see page 5). Thus, applying rule **OM15** (step 1), the minimum cardinality on the side of class PATIENT is 1, as well. Although there is no explicit structural restriction concerning the cardinalities on the role side of the class MEDICAL_TREATMENT, the analyst applied rule **OM15** | |

(step 2) and, relying on her knowledge of the field, opted for cardinalities 0:M. This derivation is depicted in Figure 13.

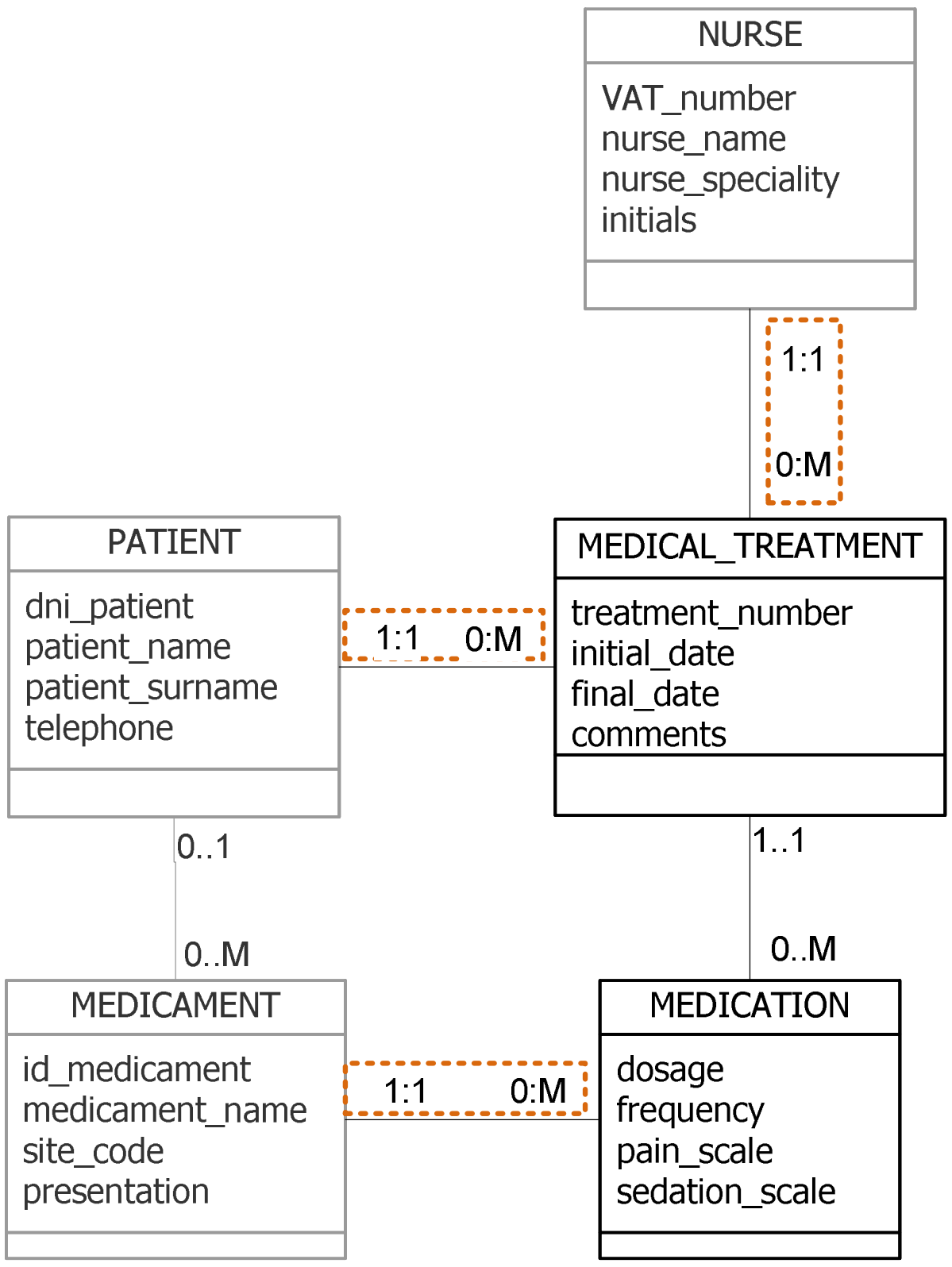

Figure 13. Specification of the cardinalities among the classes MEDICAL_TREATMENT, MEDICATION, PATIENT and MEDICAMENT

A similar reasoning applies to the cardinalities of the relationship between MEDICATION and MEDICAMENT.

Additionally, a creation service is added to each newly-derived class (see **OM18**).

| **Rule OM18.** Addition of a creation service | |
|---|---|
| **Preconditions** | |
| | Rule **OM5** has already been applied |
| **Textual explanation** | |
| 1 | Add a creation service for each newly-derived class. |
| 2 | The analyst can freely decide the name of the creation service, although it is recommended to give it the name new_class_name (i.e., the class name with the prefix new_). |
| **Example** | |
| For instance, a service named new_medical_treatment is added to the class MEDICAL_TREATMENT (see Figure 14). Similarly, the class MEDICATION is added a creation service named new_medication. | |

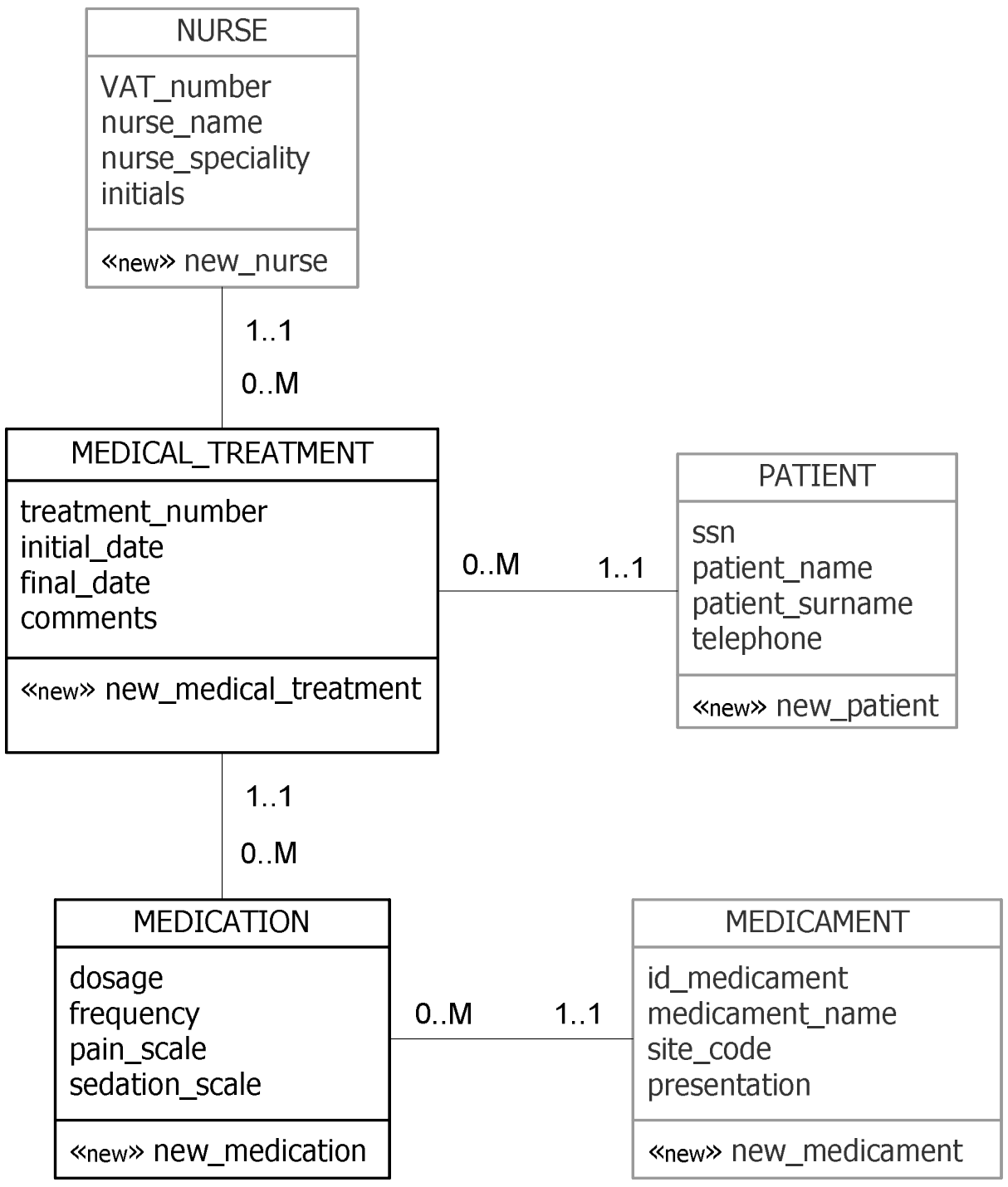

Figure 14. Addition of creation services

It is necessary to define the arguments of the creation services. To do this, the analyst uses the fields of the message structures. Data fields derive data-valued inbound arguments (see rule **OM19**), whereas reference fields derive object-valued inbound arguments (see rule **OM20**).

| **Rule OM19**. Add data-valued inbound arguments | |
|---|---|
| **Preconditions** | |
| | Rule **OM18** has already applied. |
| **Textual explanation** | |
| 1 | For each creation service, add a data-valued inbound argument for each data field contained in the complex substructure that has led to the derivation of the service. |
| 2 | The analyst can freely decide the name of the data-valued inbound argument, although it is recommended to give it the name p_atrdatafield (i. e., the data field name with the prefix p_atr) |
| 3 | The properties of the data-valued inbound arguments coincide with the properties of their corresponding attributes. This means that the data type, size and compulsory are the same that were specified when the properties of the attributes were defined. |
| **Example** | |
| For instance, the data field Treatment number has led to the derivation of the attribute treatment_number (by applying the rule **OM6**). Now it leads to the derivation of a data-valued argument named p_atrtreatment_number, which is added to the creation service new_medical_treatment.<br>The arguments and their properties are specified in Table 6. Notice that its properties are the same that its corresponding attributes (see Table 4) | |

| **Rule OM20**. Add object-valued inbound arguments | |
|---|---|
| **Preconditions** | |
| | Rule **OM19** has already been applied. |
| **Textual explanation** | |
| 1 | For each creation service, add an object-valued inbound argument for each reference field contained in the complex substructure that has led to the derivation of the service. |
| 2 | The properties of the object-valued inbound arguments are derived as follows. Take into account that the reference field has led to the derivation of a structural relationship (see rule **OM16**) and a minimum cardinality on the side of the referenced class has been defined (see rule **OM17**, step 2). |
| 3 | The argument data type is the referenced class. |
| 4 | If the minimum cardinality on the side of the referenced class is 0 then the property Null allowed of the argument is set to “yes”, else (if it is 1) the property Null allowed of the argument is set to “no”. |
| **Example** | |
| For instance, the reference field Patient has led to the derivation of a structural relationship between classes Medical_Treatment and Patient. Now it leads to the derivation of an object-valued argument named p_agrPatient, which is added to the service new_medical_treatment. This attribute defines the patient for who the medical treatment is prescribed. The data-type of the argument is the class Patient. Since the minimum cardinality of the above-mentioned structural relationship on the side of class Patient is 1 (see Figure 13) then the inbound argument does not allow null values.<br>The arguments and their properties are specified in Table 6. | |

Table 6 shows the derived inbound arguments and their properties[10].

Table 6. Specification of inbound arguments of service new_medical_treatment

| Argument name | Data type | Size | Null allowed |
|---|---|---|---|
| p_atrtreatment_number | Autonumeric | - | no |
| p_atrinitial_date | Date | - | no |
| p_atrfinal_date | Date | - | no |
| p_atrcomments | String | 200 | yes |
| p_agrPatient | Patient | - | no |

When the class diagram view corresponds to a complex business object, then apart from the creation service for each of its constituent classes another service needs to be added to the *main* class of the business object (see rule **OM21**). We refer to this kind of service as end-of-editing service because, by means of executing it, the user indicates that s/he has finished editing the message corresponding to the communicative event and that the information system reaction should be triggered.

[10] Prefixes in the argument names follow *OLIVA**NOVA*** naming conventions.

| **Rule OM21**. Add an end-of-editing service for complex business objects | |
|---|---|
| **Preconditions** | |
| | Rule **OM5** executed. |
| **Textual explanation** | |
| 1 | Add a service to each class that corresponds to a complex substructure and this corresponds to a complex business object. |
| 2 | The service name corresponds with the principal event lead a trigger event. The analyst can freely decide the name of the service, although it is recommended to give it the name event-id_name-of-the-event. |
| **Example** | |
| Therefore, a service named Treat1_prescribe_medication is added to the class MEDICAL_TREATMENT (see Figure 15). This service is triggered by the doctor whenever s/he has finished introducing the information. Only after this service is executed, the medical treatment is considered to be prescribed. | |

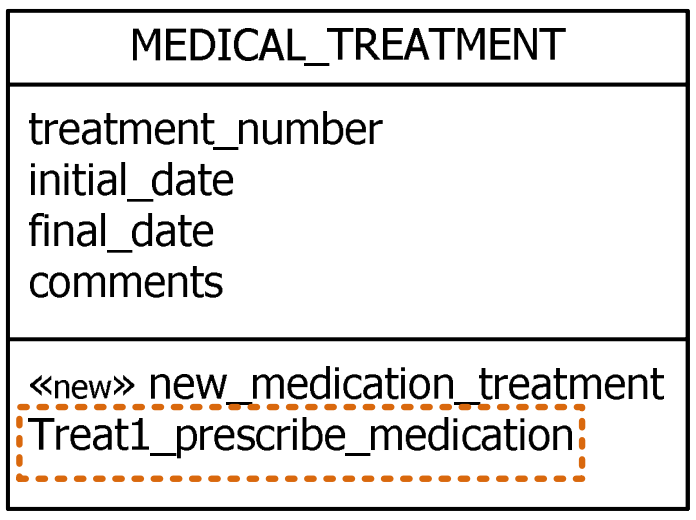

Figure 15. Class MEDICAL_TREATMENT with the service Treat1_prescribe_medication

According to the OO-Method, every service that is not a creation service needs an object-valued inbound argument that represents the instance of the class for which the service is invoked (see rule **OM22**).

| **Rule OM22**. Add inbound arguments to services that are not creation services. | |
|---|---|
| **Preconditions** | |
| | Rules **OM21** has already been applied. |
| **Textual explanation** | |
| 1 | For each service that is not a creation service, add an inbound argument that represents the instance of the class for which the service is invoked. |
| 2 | Give the argument the name p_thisClassName (i.e. the class name with the prefix p_this).<br>The argument data type is the class for which the service has been defined. |
| **Example** | |
| For instance, for the service Treat1_prescribe_medication, an inbound argument is p_thisMedicalTreatment. The argument data type is Medical_Treatment. | |

As explained above, after a doctor prescribes a medical treatment (that is, communicative event *TREAT 1*), a nurse assigns the dispensary where the treatment will be provided, and she assigns the treatment delivery date (that is, *TREAT 2*). This way, *TREAT 2* affects the same business object that is affected by the precedent event; namely, the medical treatment. For this reason, by applying rule **OM2**, the reference field Treatment belonging to the message structure of *TREAT 2* that refers to the previously-prescribed medical treatment has been marked; that is, its *Extends business object* property (column) has been set to True. Whenever an aggregation substructure includes a reference field with such a mark, then we need to extend a previously-derived class with new attributes and/or structural relationships (see rule **OM23**).

| **Rule OM23**. Extension of a class | |
|---|---|
| **Preconditions** | |
| | Rule **OM5** has already applied. |
| **Textual explanation** | |
| 1 | For each aggregation substructure that contains a reference field whose property *Extends business object* is True, extend a previously-derived class. |
| 2 | To choose the class, first select the aggregation substructure that corresponds to the business object referenced by the reference field; it will belong to a precedent communicative event (otherwise there is either an invalidity in the precedence relationships or an error has been made during the event sorting, rule **OM3**). Then obtain the class that was derived from this aggregation substructure; to do this, simply follow the traceability link. |
| 3 | Each data field that is contained in the aggregation substructure leads to the derivation of a new attribute that is added to the extended class (see rule **OM6**). |
| 4 | Each reference field that is contained in the complex substructure and is not marked leads to the derivation of a new structural relationship between the class being extended and the class that corresponds to the reference field (the derivation of structural relationships and their cardinalities is explained in rules **OM16** and **OM17**). |
| 5 | In any case, take into account that the minimum cardinality in the side of the referenced class is necessarily 0. The moment the instance is created this link is not created; it occurs later). |
| **Example** | |
| Figure 16 shows the specification of the message structure of communicative event TREAT 2 . The reference field Treatment refers to a medical treatment and, since its property *Extends business object* is True, then the class that corresponds to this business object is extended; that is, MEDICAL_TREATMENT (Figure 16 depicts how to select the class).<br><br>The data fields in the message structure lead to adding new attributes to the class (according to rule **OM6**); this way, the field delivery_date leads to adding an attribute named delivery_date to the class MEDICAL_TREATMENT. In turn, the reference fields lead to adding new structural relationships between the class and previously-derived classes; this way, the reference field Dispensary leads to adding a structural relationship between MEDICAL_TREATMENT and DISPENSARY (according to rule **OM16**). | |

OM Rule 23

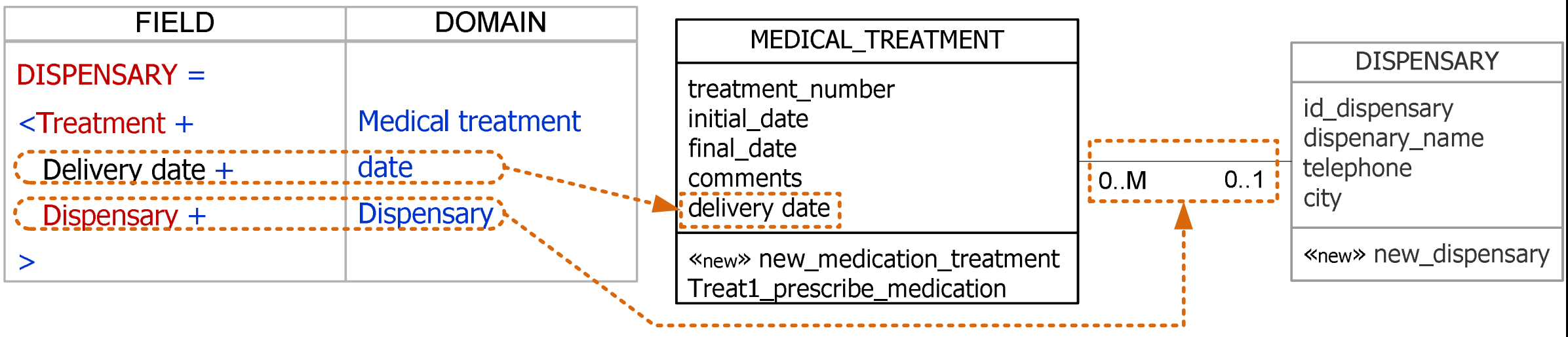

Figure 16. Example of a class extension

The maximum cardinality in the side of DISPENSARY is 1 (according to rule **OM17**); the minimum cardinality is necessarily 0 because medical treatments are not assigned to dispensaries the moment they are prescribed, but this assignment occurs later in time. For the same reason, the relationship is dynamic.

All the attributes that are added to a class as a result of a class extension should have their properties defined as explained in rule **OM24**.

| **Rule OM24.** Attribute properties of extended classes | |
|---|---|
| **Preconditions** | |
| | Rule **OM23** has already been applied. |
| **Textual explanation** | |
| 1 | The attributes are not part of the identification function. |
| 2 | The attribute type is Variable. |
| 3 | The attributes are not requested upon creation. |
| 4 | The attributes allow nulls. |
| 5 | The attributes data types are derived from the domain of the field (according to rule **OM10**). |
| **Example** | |
| According to the textual explanation, the properties of the attribute delivery_date are showed in Table 7. | |

Table 7. Properties of the attribute delivery_date

| **Attr. name** | **Id** | **Attr. type** | **Data type** | **Size** | **Requested** | **Null allowed** |
|---|---|---|---|---|---|---|
| delivery_date | no | Variable | date | | no | yes |

Whenever new attributes are added to an extended class, a service needs to be added in order to set their values (see rule **OM25**). According to the OO-Method, these attributes are said to be *own events* (a.k.a. edit events), what means that they are neither creation events nor destruction events.

| **Rule OM25**. Add services to extended classes | |
|---|---|
| **Preconditions** | |
| | Rule **OM24** has already been applied. |
| **Textual explanation** | |
| 1 | Add a service to the extended class in order to allow setting the values of the newly-added attributes. |
| 2 | If the class has been extended with just one attribute, then the service can be named set_attribute_name (i.e. the name of the attribute with the prefix set_). Else, select an appropriate name that represents all the information being provided. |
| 4 | For each structural relationship that has been derived during the extension of the class, add two shared services to the involved classes intended to add and to destroy links between instances (ins_class_name and del_class_name, respectively). |
| **Example** | |
| A service named set_delivery_date is added to the class to set the value for the newly-added attribute delivery_date. Furthermore, given the cardinality of the structural relationship, two shared services are included in both classes (this is prescribed by the OO-Method [7]); namely an insertion shared service named ins_ dispensary and a deletion shared service named del_dispensary. | |

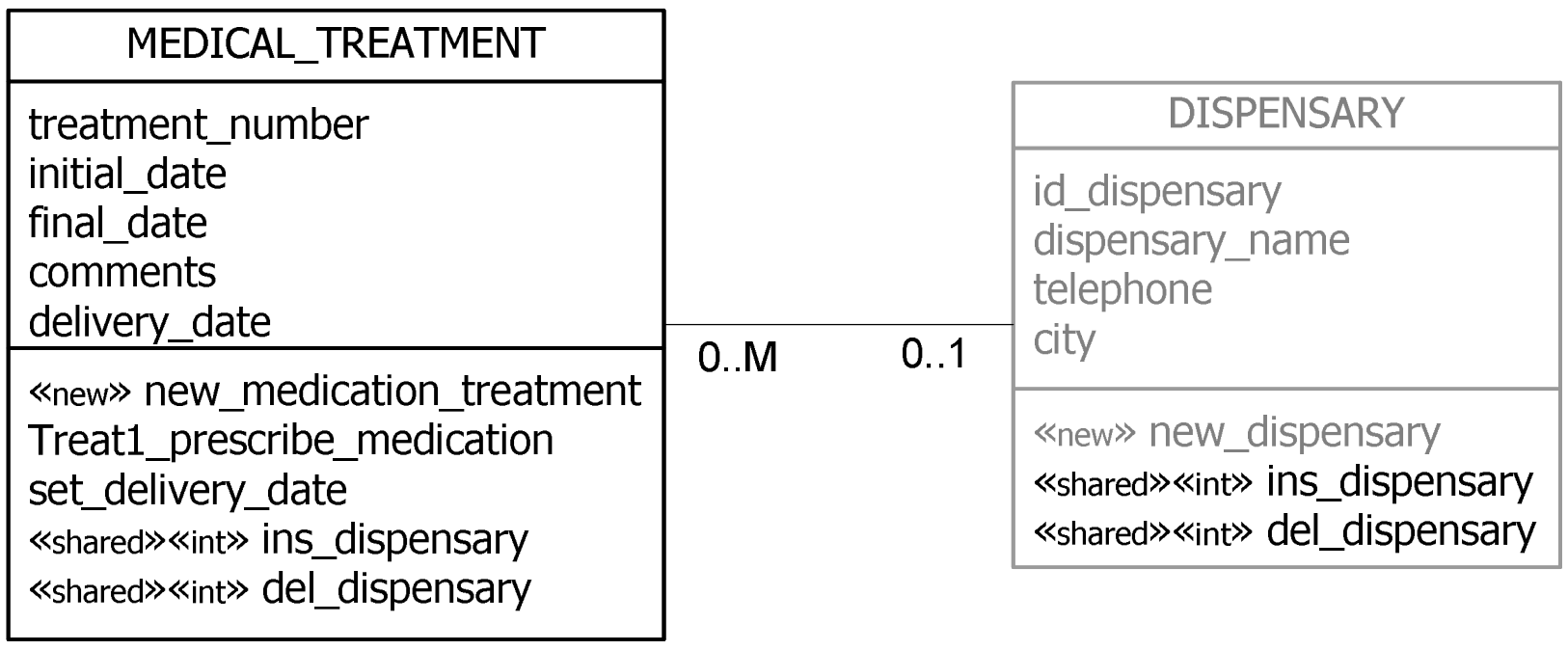

Figure 17. Addition of set services and shared services to the classes MEDICAL_TREATMENT and DISPENSARY

When the reaction to a communicative event requires the execution of several services of the same class, then a transaction is created in order to ensure their atomic execution (see rule **OM26**). The rationale for creating such a transaction is that it is not a good practice to delegate to the user of the software application the responsibility of executing all the needed services one after another; it is safer to encapsulate all of them in a transaction.

| **Rule OM26**. Creation of transactions | |
|---|---|
| **Preconditions** | |
| | Rule **OM25** has already been applied. |
| **Textual explanation** | |
| 1 | Create a transaction when is necessary to execute several services atomically. Whenever several services are defined for a class as a result of the processing of a single communicative event (they will often be an event to set the value of attributes and one or several insertion shared events to create links with other instances), then a transaction needs to |
| | be created to ensure the atomic execution of all the services). |
| 2 | Set all the services involved in the transaction as internal. The transaction needs to be the only service made available for the user. |

**Example**

For instance, ASSIGN_DISPENSARY is defined in order to execute atomically the events set_delivery_date and ins_dispensary (see Figure 18).

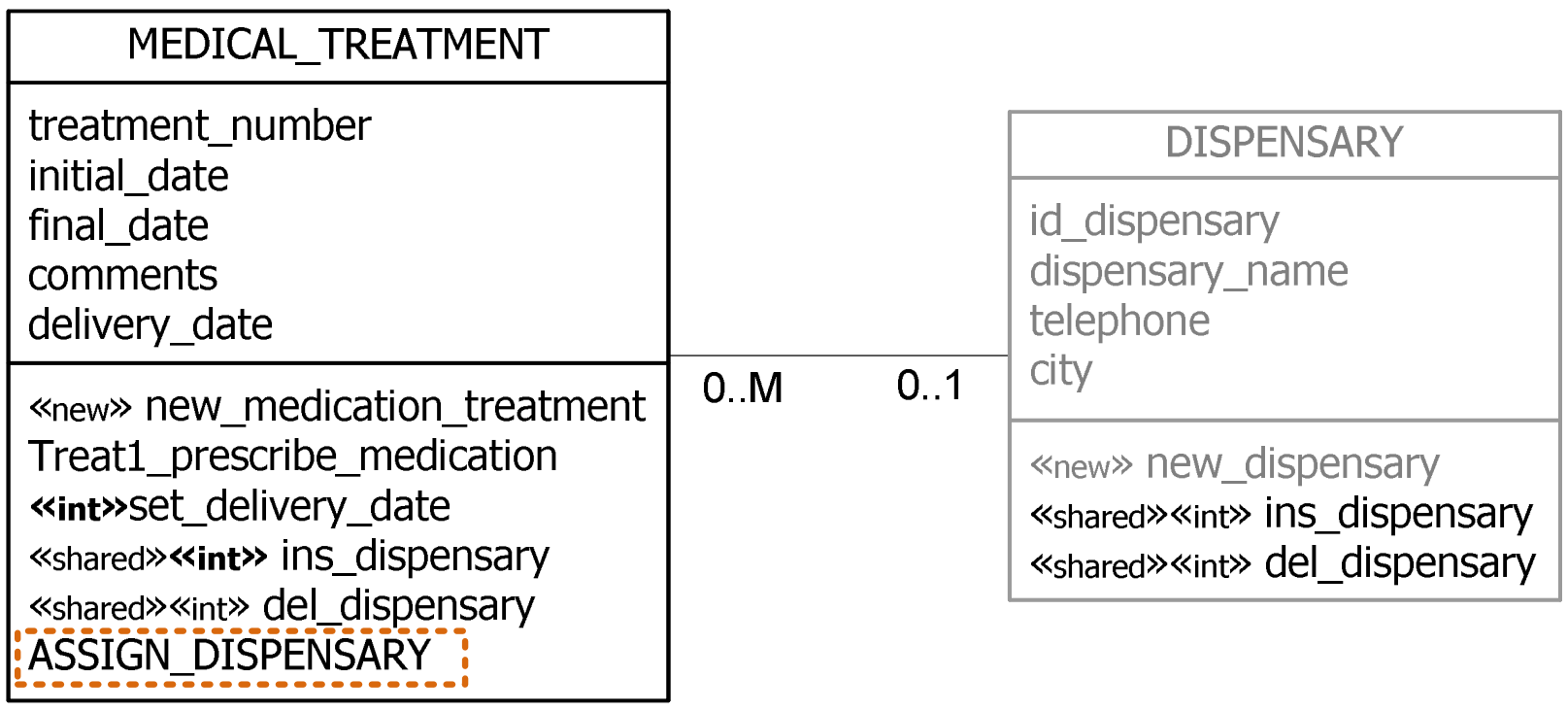

Figure 18. Creation of the transaction ASSIGN_DISPENSARY that corresponds to the reaction to communicative event *TREAT 2*

Figure 19 shows the class diagram (Object Model) that results from processing the events Treat 1, Treat 2 and its predecessors

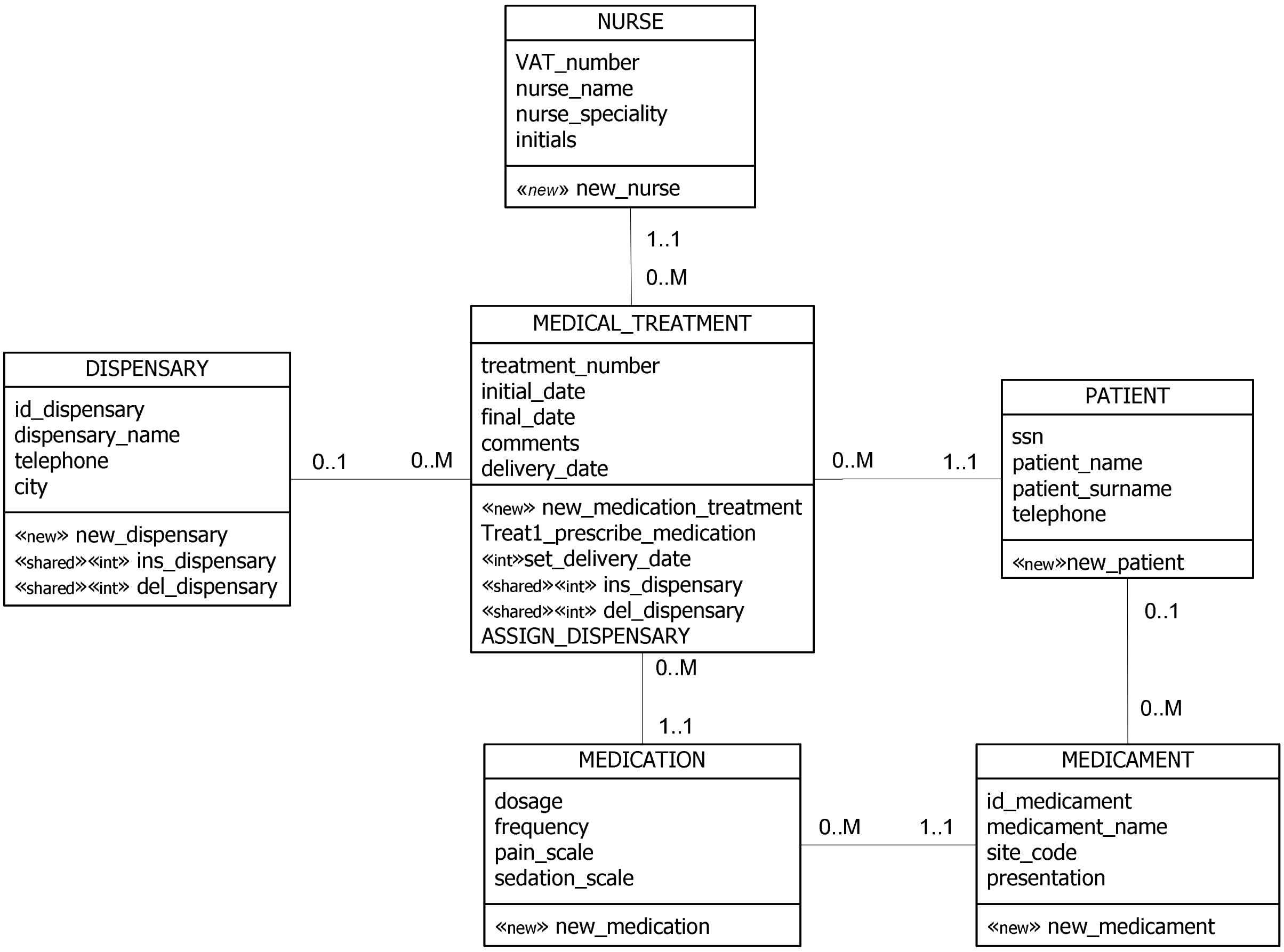

Figure 19. Class diagram obtained processing all the events in the example.

# 3. DERIVATION OF THE DYNAMIC MODEL

## 3.1. Derivation Strategy

We first present an overview of the derivation strategy (see Figure 1); with regards to the Dynamic Model, for each class of an Object Model, a state transition diagram (STD) is derived. The main input artefact for this derivation is the Communicative Event Diagram (See Figure 20.b). Roughly speaking communicative events are converted into transitions (e.g. IV, II) and precedence relationships are converted into states (e.g. IVED, IIED). Additional transitions can be added to the state transition diagram; for instance, transitions that correspond to edition and destruction atomic services, as well as transitions that corresponds to services that take part in a transaction.

To increase the comprehensibility of the examples, precedence relations and transitions have been labelled with a name that is derived from the name of the precedent event. The English participle suffix *–ed* is added to the event name. This intuitively represents the state in which an object remains after an event. For instance, after an occurrence of the event IV the affected object remains in the state IVED.

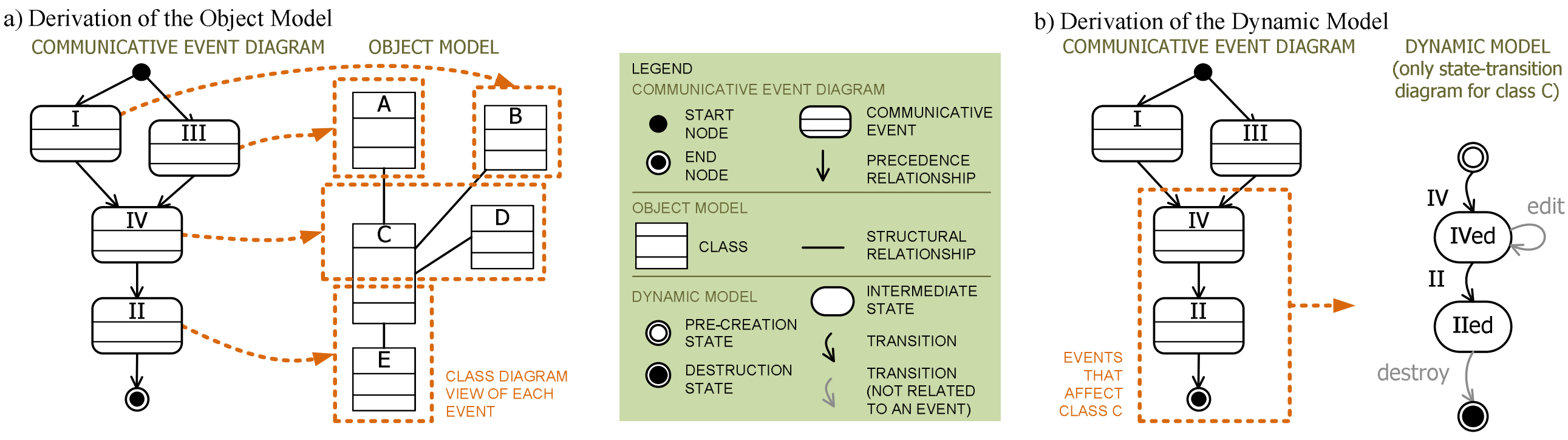

Figure 20. Simplified example of the derivation of the structural and behavioural views of the conceptual model

## 3.2. Derivation Guidelines

The first step is to obtain, for each class in the Object Model, a sub-diagram that contains only the communicative events that affect the given class is obtained (see Rule **DM1**). The derivation of the Object Model from the requirements model also produces traceability mappings that specify which classes have been created or extended by which communicative events; thus, obtaining the sub-diagram is straightforward.

| **Rule DM1.** Production of the sub-diagram for class C | |
|---|---|
| **Preconditions** | |
| | None |
| **Textual explanation** | |
| 1 | Duplicate the communicative event diagram. |
| 2 | Remove every communicative event for which there is no traceability mapping to class C. |
| 3 | Remove any loopbacks (if any). |
| 4 | Add a start node to the CED (if it does not already exist). |
| 5 | For each initial communicative event, add a precedence relation that departs from the start node and arrives at the event (if the relation |
| | does not already exist). (End nodes and their corresponding precedence relations are not needed for derivation purposes, but they do not need to be removed.) |

**Example**

Following the previous example (see Figure 1), the communicative events that affect a class C are identified and the sub-diagram is produced.

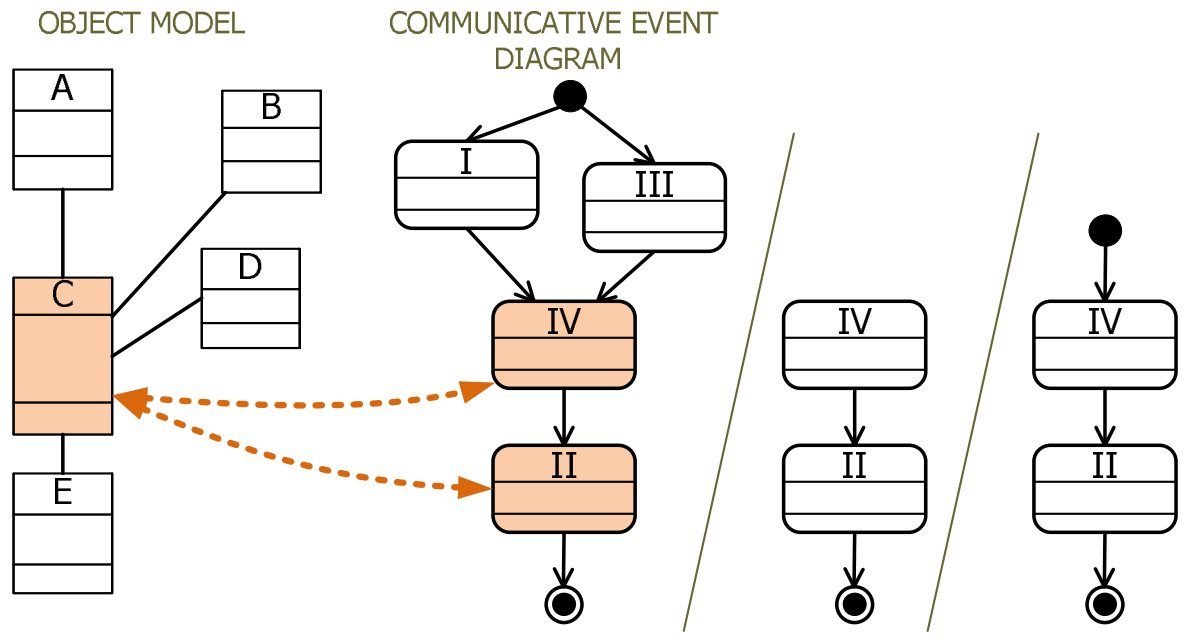

Figure 21. Sub-diagram corresponding to class C (from example in Figure 1)

Then an empty STD is created and a pre-creation state is added (see rule **DM2**).

| **Rule DM2.** Initialisation of the STD | |
|---|---|
| **Preconditions** | |
| 1 | The CED has a start node (referred as n). |
| **Textual explanation** | |
| 1 | Create an empty STD. |
| 2 | Add a pre-creation state to the STD. |
| 3 | Add a traceability link between the start node and the newly-created state. |
| **Pseudocode** | |
| 1 | std = std.new() |
| 2 | s = state.new()<br>s.name=“Pre_creation”<br>s.type=“pre_creation” |
| 3 | n.trace_state.add(s) |
| **Example** | |

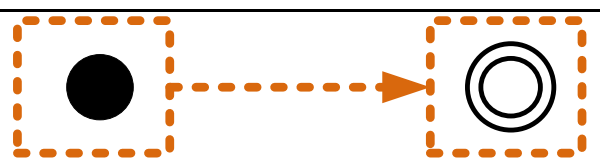

Figure 22. Examples of transformation of a start node

Rules **DM2**, **DM3** and **DM4** are the main derivation guidelines that produce intermediate states and transitions. Rule **DM2** accounts for communicative events that are not specialised and that do not have and-joins in their incoming precedence relations.

| | |
|---|---|
| **Rule DM2.** Transformation of a communicative event A | |
| **Preconditions** | |
| 1 | All the precedent events of e A have already been processed (For all pe in precedents(e), Exists t in event2state, t.event=pe) |
| 2 | A has only one precedent event or all the precedences are composed in an or-merge. |
| **Textual explanation** | |
| 1 | Add a state to the STD and label it with the identifier of the event plus the suffix -ed (e.g. given an event identifier A, the state label is Aed). |
| 2 | Else, for each communicative event P that is a precedent of A, add a transition to the STD; the transition departs from the state that is traced to the precedent event and it arrives to the newly-added state. |
| 3 | Map the newly-added transitions to the class service that corresponds to the communicative event A. The service name should appear in the transition label. |
| 4 | Add the names of the agent classes that correspond to the support actor of A to the list of agents of the newly-added transitions. |
| 5 | Add a traceability link between the communicative event and the newly-added transitions as well as between the communicative event and the newly-created state. |
| **Pseudocode** | |
| 1 | s = state.new() |
| | s.name= e.name+“ed” |
| | s.type=“intermediate” |
| 2 | For each pe in e.precedents do |
| |   t = transition.new() |
| |   t.initial_state= pe.trace_state -- podría haber varios! |
| |   t.end_state=s |
| 3 |   t.service=e.trace_service -- podría haber varios! |
| |   -- e.trace_service was linked while deriving the Object Model |
| |   -- (poner como pies de página, uno para todos los casos?) |
| 4 |   For each ag in e.supports.actor.trace_agent_class |
| |     t.agent_class.add(ag) |
| 5 |   e.trace_transition.add(t) |
| | e.trace_state.add(s) |
| **Example** | |

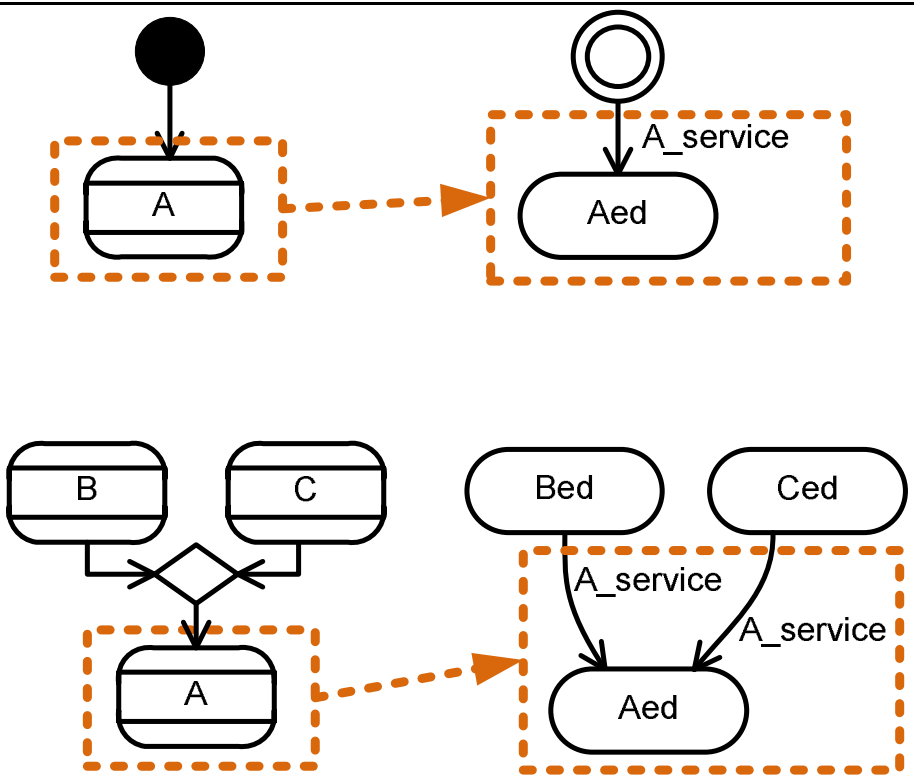

Figure 23. Examples of transformation of a communicative event

Rule **DM3** accounts for specialised communicative events. This leads to different paths in the STD.

| **Rule DM3.** Transformation of a specialized communicative event A **e** | |
|---|---|
| **Preconditions** | |
| 1 | All the precedent events of communicative event A have already been processed. |
| 2 | A only has one precedent event or all the precedences are composed in an or-merge. |
| **Textual explanation** | |
| 1 | For each event variant of the specialized communicative event, add a state to the STD and label it with the identifier of the variant plus the suffix -ed (e.g. A1ed, A2ed). |
| 2 | Add a traceability link between |
| 3 | For each communicative event P that is a precedent of A do |
| | For each event variant Ai of the specialized communicative event, add a transition to the STD; the transition departs from the state that corresponds to the precedent communicative event (or to the pre-creation state, in case A is an initial event) and it arrives at the state that corresponds to the variant. |
| 4 | Map the newly-added transitions to the class service that corresponds to the communicative event A. Also, add a control condition to each transition; the condition is derived from the specialisation condition that corresponds to the event variant (instead of message structure fields, their corresponding class arguments are used). |
| | Add a default message to each newly-added transitions (this message will be displayed to the user if the transition cannot be executed. |
| 5 | Both the service name and the condition should appear in the transition label, as prescribed by the OO-Method: service_name *when* condition |
| 6 | Add the names of the agent classes that correspond to the support actor of A to the list of agents of the newly-added transitions. |
| 7 | Add a traceability link between the event variants and the newly-added transitions, and between the event variants and their corresponding, |
| 8 | newly-created states. Idem for the specialised communicative event. |

**Pseudocode**

| | |
|---|---|
| 1 | For each v in e.variants do |
| | s=state.new() |
| | s.name= v.name+"ed" |
| | s.type="intermediate" |
| 2 | e.trace_state.add(s) |
| 3 | For each pe in e.precedents do |
| | For each v in e.variants do |
| | t = transition.new() |
| | t.initial_state= pe.trace_state -- podría haber varios! |
| | t.end_state=v.trace_state -- idem |
| 4 | t.service=e.trace_service -- idem |
| | -- este es solucionable si se diferencia el servicio de fin de edición |
| | c=control_condition.new() |
| | c.formula=from_field_to_attributes(v.condition) -- definirla inform. |
| 5 | t.message="This action cannot be executed" -- default message |
| 6 | -- displaying the transition label is a duty of the diagramming tool |
| 7 | For each ag in e.supports.actor.trace_agent_class |
| | t.agent_class.add(ag) |
| 8 | v.trace_transition.add(t) |
| | v.trace_state.add(s) |
| | e. trace_transition.add(t) |
| | e.trace_state.add(s) |

**Example**

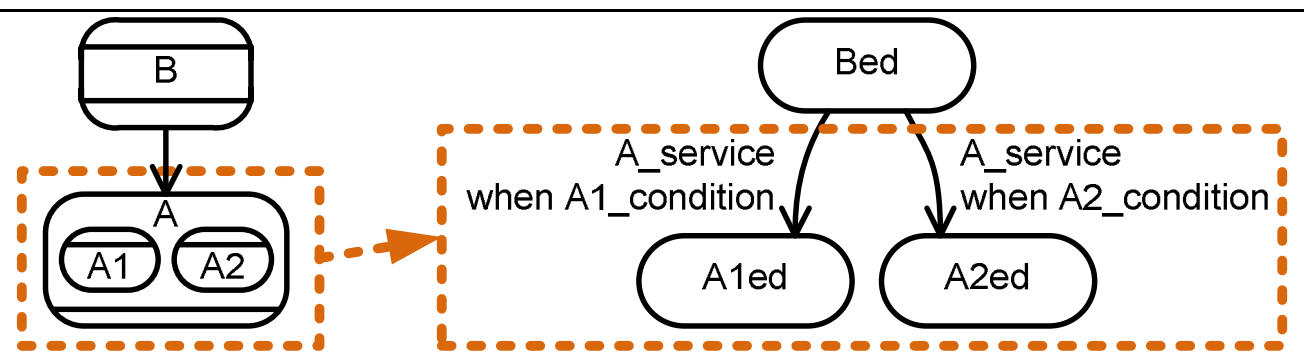

Figure 24. Example of transformation of a specialised communicative event

Rule **DM4** accounts for and-joins. This implies the creation of auxiliary states that constitute a logical network.

| **Rule DM4.** Transformation of an and-join over a communicative event A | |
|---|---|
| **Preconditions** | |
| 1 | All the precedent events of A have already been processed. |
| 2 | A has two or more precedent events and all the precedences are composed in an and-join. |

| **Textual explanation** | |
|---|---|
| 1 | Add a state to the STD and label it with the identifier of the communicative event plus the suffix -ed (e.g. Aed). |
| | Then a logical network is created and later simplified as follows. |
| 2 | All the allowed sequences of events are enumerated as columns in a matrix. In each row a communicative event is appended in a way that each matrix cell contains the set of events that have occurred in that particular sequence. |
| | Links from one cell to another indicate the event that occurs. |
| 3 | The matrix is simplified by joining those sets that are equal (i.e. disregarding the order in which the events occur). The corresponding links are redirected to the new destination. The last row will always be simplified to a single cell. |
| 4 | For each cell in the matrix derive one state in the STD (if the state does not exist yet). Cells that have been joined lead to adding auxiliary states to the STD that support the logical network. |
| | For each links between cells in the matrix derive a transition in the STD (if the transition does not exist yet). |
| 5 | Map each transition to the class service that corresponds to the communicative event in the link from where it has been derived. The service name should appear in the transition label. |
| 6 | For each transition, identify the support actors of the event in the link from where it has been derived. |
| 7 | Add the names of the agent classes that correspond to the support actors to the list of agents of the transition. |
| 8 | Add a transition to the STD; the transition departs from the last state of the logical network and it arrives at the state that has been created in step 1 of this rule. |
| 9 | |

**Example**

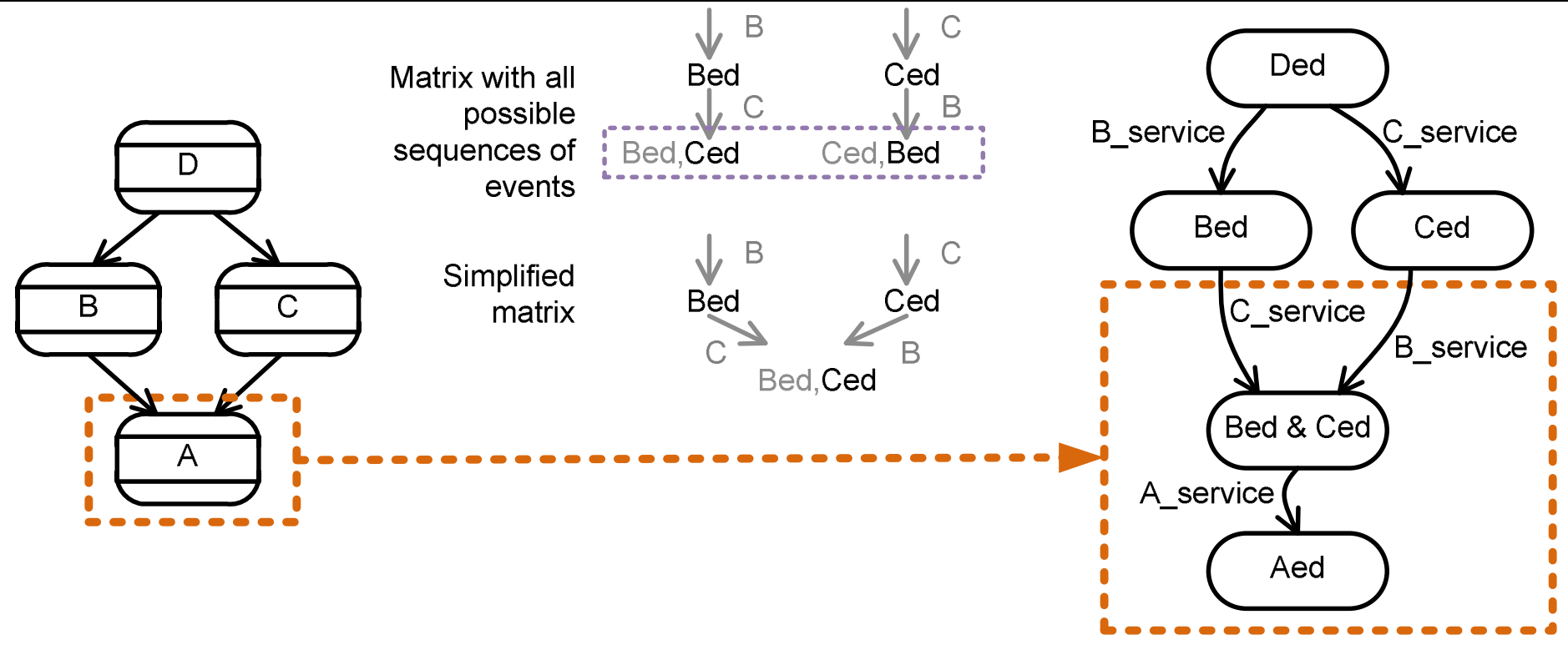

Figure 25. Example of transformation of an and-join